\documentclass[twocolumn,showpacs,preprintnumbers,prd,superscriptaddress,nofootinbib]{revtex4}
\usepackage{epsfig}
\usepackage{graphicx}
\usepackage{amsmath,amssymb,amsfonts}
\usepackage{array}
\usepackage{url}
\usepackage{hyperref}
\usepackage{lineno}
\usepackage{xspace}
\usepackage[usenames,dvipsnames]{color}

\begin{document}
\def\jour#1#2#3#4{{#1} {\bf#2}, #4 (19#3)}
\def\jou2#1#2#3#4{{#1} {\bf#2}, #4 (20#3)}
\def\PRL{{Phys. Rev. Lett. }}
\def\EPJA{{Eur. Phys. J.  A }}
\def\EPL{{Europhys. Lett.}}
\def\PRv{{Phys. Rev. }}
\def\PRC{{Phys. Rev.  C }}
\def\PRD{{Phys. Rev.  D }}
\def\JAP{{J. Appl. Phys. }}
\def\AJP{{Am. J. Phys. }}
\def\NIMA{{Nucl. Instr. and Meth. Phys. A }}
\def\NPA{{Nucl. Phys. A }}
\def\NPB{{Nucl. Phys.  B }}
\def\NPBP{{Nucl. Phys.  B (Proc. Suppl.) }}
\def\NJP{{New J.  Phys. }}
\def\EPJC{{Eur. Phys. J.  C }}
\def\PLB{{Phys. Lett. B }}
\def\PHY{{Physics }}
\def\MPLA{{Mod. Phys. Lett. A }}
\def\PRp{{Phys. Rep. }}
\def\ZPC{{Z. Phys. C }}
\def\ZPA{{Z. Phys. A }}
\def\PPNP{{Prog. Part. Nucl. Phys. }}
\def\JPG{{J. Phys. G }}
\def\CPC{{Comput. Phys. Commun. }}
\def\APP{{Acta Physica Pol. B }}
\def\AIP{{AIP Conf. Proc. }}
\def\JHEP{{J. High Energy Phys. }}
\def\PSC{{Prog. Sci. Culture }}
\def\NCim{{Nuovo Cim. }}
\def\SNC{{Suppl. Nuovo Cim. }}
\def\SJNP{{Sov. J. Nucl. Phys. }}
\def\SPJ{{Sov. Phys. JETP }}
\def\JLet{{JETP Lett.}}
\def\PTP{{Prog. Theor. Phys. }}
\def\PTPS{{Prog. Theor. Phys. Suppl. }}
\def\IANSF{{Izv. Akad. Nauk: Ser. Fiz. }}
\def\JPCS{{J. Phys. Conf. Ser. }}
\def\AHEP{{Adv. High Energy Phys. }}
\def\IJMPE{{Int.  J. Mod. Phys. E }}
\def\ARNPS{{Ann. Rev. Nucl. Part. Sci. }}

\def\ct{\cite}
\def\sNN{\sqrt{s_{N\!N}}}
\def\sNNq{s_{N\!N}}
\def\sppq{s_{pp}}
\def\spp{\sqrt{s_{pp}}}
\def\eNN{\varepsilon_{N\!N}}
\def\pbp{{\bar p}p}

\def\col{Collab.}
\def\bi{\bibitem}
\def\ea{{\sl et al.}}
\def\eg{{\sl e.g.}}
\def\vrs{{vs}}
\def\ie{{i.e.}}
\def\va{{\sl via}}
\def\nopar{\noindent}
\def\bi{\bibitem} 
\def\lssim{\stackrel{<}{_\sim}}
\def\gtsim{\stackrel{>}{_\sim}}

\title{ Multihadron production dynamics exploring the energy balance \\ in
hadronic and nuclear collisions
}
\author{Edward K. G. Sarkisyan}
\email {sedward@cern.ch}
\affiliation{Experimental Physics Department, CERN, 1211 Geneva 23, Switzerland}
\affiliation{Department of Physics, The University of Texas at
Arlington, Arlington, TX 76019, USA}
\author{Aditya Nath Mishra}
\email {Aditya.Nath.Mishra@cern.ch}
\affiliation{Discipline of Physics, School of Basic Sciences, Indian
Institute of Technology Indore,  Indore 452020, India}
\author{Raghunath Sahoo}
\email{Raghunath.Sahoo@cern.ch}
\affiliation{Discipline of Physics, School of Basic Sciences, Indian
Institute of Technology Indore,  Indore 452020, India}
\author{Alexander S. Sakharov}
\email{Alexandre.Sakharov@cern.ch}
\affiliation{Department of Physics, CERN, 1211 Geneva 23, Switzerland}
\affiliation{Department of Physics, New York University, New York, NY
10003, USA}
\affiliation{Physics Department, Manhattan College, Riverdale, NY 10471,
USA}

\begin{abstract}
The relation of multihadron production in nucleus-nucleus and
(anti)proton-proton collisions is studied by exploring the
collision-energy and centrality dependencies of the charged particle mean
multiplicity in the measurements to date. The study is performed in the
framework of the recently proposed effective-energy approach which, under
the proper scaling of the collision energy, combines the constituent quark
picture with Landau relativistic hydrodynamics counting for the
centrality-defined effective energy of participants. Within this approach,
the multiplicity energy dependence and the pseudorapidity spectra from the
most central nuclear collisions are well reproduced. The study of the
multiplicity centrality dependence reveals a new scaling between the
measured pseudorapidity spectra and the calculations. By means of this
scaling, referred to as energy-balanced limiting fragmentation scaling,
one reproduces the pseudorapidity spectra for all centralities. The
scaling elucidates some differences in the multiplicity and midrapidity
density centrality dependence obtained at RHIC and LHC. These findings
reveal an inherent similarity in the multiplicity energy dependence from
the most central collisions and centrality data. Predictions are made for
the mean multiplicities to be measured in proton-proton and heavy-ion
collisions at the LHC.
 %end of abstract

\end{abstract}

\maketitle
 %\nopar
 \section{Introduction}
 \label{sec:intro}
 %{\bf 1. } 
 Study of global observables of multiparticle production and their   
universality in different types of 
high-energy 
collisions is  
 of a crucial 
importance 
 for  
 understanding 
 the   
underlying dynamics 
 of strong interactions. 
 Recently, 
the universality of multiparticle production in nucleus-nucleus 
and
hadron-hadron
collisions has been reported 
 exploiting
concept of 
 centrality-defined 
   effective energy \ct{edward2}
 employed 
 for the 
data
interpreted 
 in terms of the 
approach of the 
dissipating energy 
of quark participants 
\ct{edward,edwarda}. 
 This approach
combines the constituent
quark picture together with Landau relativistic hydrodynamics
  and  interrelates 
 multihadron production in 
 different types of 
 collisions.
 Within this 
 % The 
 picture 
 % of the dissipating energy of participants 
 % allows 
one can 
 % as well  
 % to 
successfully explain \ct{edward,edwarda}   
 the 
 %earlier observed \ct{lep}
 scaling between the 
charged particle mean multiplicity in 
 $e^+e^-$ and $pp/\pbp$ collisions 
 \ct{lep} 
 and 
 the 
 universality 
 of both  the  multiplicity and 
 the 
 midrapidity 
 pseudorapidity density 
 measured
 in the 
most central 
nuclear 
collisions and  in  $e^+e^-$ annihilation \ct{phobos-sim}.\footnote{Elsewhere 
in this paper, the  multiplicity is 
defined as 
being averaged over events in 
what is referred to as the ``mean'' 
multiplicity. No other averaging, \eg\ over centralities in nuclear data, 
is considered.
} 
The universality of the multihadron production irrespective 
 of the collision species, 
 an intrinsic
 feature 
 of the 
 %participant 
 dissipating energy approach,
  is widely discussed 
 nowadays
 \ct{book,pp-mult-rev,pdg14}. 

In this paper, in the framework of the 
approach
of the dissipating effective energy 
of constituent quark participants, or, for brevity, the 
 participant dissipating energy (PDE)  
approach,   we extend the 
previous 
studies
of the charged particle mean multiplicity 
\ct{edward,edwarda} 
 %spreading
 %the c.m. energy range of 
 %nucleus-nucleus collisions up 
to 
 %the 
 LHC 
energies. 
We show that the multiplicity energy dependence of head-on collisions is 
well described within the proposed approach.   
In addition, here we study  the dependence of the multiplicity 
on the 
number of (nucleon)
participants at RHIC and LHC. 
We introduce a new scaling, 
 referred to as 
 %called 
 the energy-balanced  limiting 
fragmentation
scaling, which allows 
 us
 to describe the pseudorapidity density spectra  
independently of the centrality of collisions. 
 Using this scaling, a complementarity between  the multiplicities 
measured in head-on
nuclear collisions and  obtained from the centrality data is 
 found. The 
study clarifies 
some differences 
 of the  
centrality dependence 
of  multiplicities 
measured at
RHIC and LHC. Finally, 
 predictions are made for 
 the  
 charged particle mean multiplicities  
in $pp$ and 
 heavy-ion collisions at 
 the
LHC.
\\

 \vspace*{-.45cm}
 % \nopar
 \section{The participant dissipating energy approach}
 %{\bf 2. } 
 \label{sec:pdea}
In this section, we briefly describe the 
 PDE
approach, 
 %of the dissipating effective energy 
 %of constituent quark participants, 
as 
 it is 
proposed in \ct{edward2,edward,edwarda}. 
This approach 
quantifies the process of particle production in terms of the 
amount of energy
deposited by interacting constituent quark participants inside the small
Lorentz-contracted volume formed at the early stage of a collision.
The whole process of a collision  
is then 
 represented 
 as the expansion of an initial state 
and the subsequent breakup into 
particles.  
This approach resembles the Landau 
phenomenological hydrodynamic approach  of multiparticle 
production in 
relativistic particle interactions \ct{Landau}, which was found to be in a 
good 
agreement with the multiplicity data in particle and nuclear collisions in 
the wide energy range \ct{Landau-difint}. In 
the 
 picture
considered
 here, 
 the Landau hydrodynamics  is 
 employed 
 in the framework of 
 constituent 
(or
dressed) 
quarks, in accordance with the additive quark 
 model 
 \ct{cqm-LevFr,cqm-Lip,cqm-KokVH,cqm-KokBook};
 for recent 
comprehensive review on 
soft hadron interactions in the additive quark model, see
\ct{constitq}.
 This 
 means
 %%makes 
 the secondary particle  production 
 %%to be  
 is
 basically driven by 
 the amount of the initial 
 {\it effective} energy 
 deposited 
 by
 constituent 
quarks 
 into the Lorentz-contracted 
 %overlap 
 region. In 
 $pp/\pbp$ collisions, a single constituent 
 quark from each nucleon 
 is considered to 
take part in a collision, and 
 the remaining quarks
 %rest 
 are 
treated as spectators. 
 The spectator quarks do not participate in the 
secondary particle 
production, but they  
result 
 in  %%to 
 a 
 formation
of leading particles and 
carry away a 
significant part of the collision energy.
    Thus, the effective energy for the production of secondary 
particles is the energy of interaction of  a single quark pair, \ie\ 
1/3 of 
the entire 
nucleon energy. 
 On the 
 contrary,
 in the head-on heavy-ion collisions, 
 the participating nucleons 
 are considered colliding 
 %by 
 with
 all three 
constituent quarks from each nucleon. 
 %which 
 This
 makes  
 the 
 whole 
 energy of the colliding nucleons (participants) 
 available for 
 the 
 secondary
 particle production. 
 Within this picture,
 one 
 expects 
 %that 
 the results for 
bulk 
 observables 
 %measured  in 
 from
 head-on
 heavy-ion collisions at 
the c.m. energy per nucleon, $\sqrt{s_{NN}}$,
to be 
similar to those from 
 the 
 $pp/\pbp$ collisions but 
 %at 
   corresponding to
 a 
 three times larger 
 c.m. energy, \ie\ at $\sqrt{s_{pp}} \simeq 3\sqrt{s_{NN}}$. 
 Such 
 %%an 
 a
 universality is found to correctly predict \ct{edward} the 
value 
of the midrapidity density in $pp$ interactions measured at the  
TeV LHC energies \ct{CMSwe}. In addition, the multiplicity measurements 
in $pp/\pbp$ interactions up to TeV energies are shown 
to 
be well 
reproduced
 by 
 $e^+e^-$ data as soon as the inelasticity is set to 
 $\approx$0.35 \ct{pp-mult-rev}, \ie\  effectively
 1/3 of the  hadronic 
interaction energy.
 This is
 in 
agreement with the dissipation energy picture where the 
structureless colliding leptons are considered to deposit 
their
 total energy into the Lorentz-contracted volume, similarly
 to nucleons in 
head-on nuclear collisions \ct{edward}. 
 For recent discussion on the universality of hadroproduction up to LHC 
energies, see  
\ct{pdg14}. 

Combining the above-discussed ingredients of the constituent quark 
picture 
and  Landau hydrodynamics, one obtains 
 the relationship 
 between
  charged particle rapidity 
 density per participant 
 pair, $\rho(\eta)=(2/N_{\rm{part}})dN_{\rm{ch}}/d\eta$ at midrapidity 
 ($\it{\eta} \approx $~0), in heavy-ion collisions 
 and 
 in $pp/\pbp$ collisions: 

  %Eqn.1
 \begin{equation}
 \frac{\rho(0)}{\rho_{pp}(0)} = 
 \frac{2N_{\rm{ch}}}{N_{\rm{part}}\, N_{\rm{ch}}^{pp}} 
 \, \sqrt{\frac{L_{pp}}{L_{N\!N}}} \, ,\:\:\:\:\:
 \spp=3\sNN\, .
 \label{eqn1}
 \end{equation}
  In Eq.(\ref{eqn1}), 
 %the Gaussian form of 
 the relation of the pseudorapidity 
 density and the mean multiplicity 
 is applied in its Gaussian form as obtained in Landau hydrodynamics.
 The 
 factor $L$ 
 is 
 defined 
 as 
  $L =  
 {\ln}({\sqrt{s}}/{2m})$.
  According to the 
 approach considered, 
 $m$ 
 is 
 the proton mass, $m_{p}$, in nucleus-nucleus 
collisions 
 and 
 the constituent quark mass in $pp/\pbp$ collisions 
 %which is 
 set 
to 
 $\frac{1}{3}m_{{p}}$.
 $N_{\rm{ch}}$ and 
 $N_{\rm{ch}}^{pp}$ 
are the mean multiplicities in nucleus-nucleus and nucleon-nucleon 
collisions, respectively,  
and
 $N_{\rm {part}}$ is the number of participants. 

 Solving 
 Eq.~(\ref{eqn1}) 
 for
the multiplicity $N_{\rm{ch}}$ 
 at 
 a
 given
rapidity 
 density $\rho(0)$ 
 at $\sqrt{s_{NN}}$,
and 
for
the rapidity 
density $\rho_{pp}(0)$ and the multiplicity $N_{\rm{ch}}^{pp}$ at $3 
\sqrt{s_{NN}}$, 
 one finds:

\begin{eqnarray}
\nonumber
& & 
\frac{2N_{\rm{ch}}}{N_{\rm{part}}}  =
  N^{pp}_{\rm ch} \,
   \frac{\rho(0)}{\rho_{pp}(0)} \,
  \sqrt{1-\frac{2 \ln 3}{\ln\, (4.5 \sNN/m_ p)}}\,,  \\
& &    \sNN=\spp/3 \, .  
  \label{pmult} 
\end{eqnarray}
 \nopar

Further development, 
as outlined below, treats this 
dependence in
 terms of centrality \ct{edward2}. 
 The centrality is 
 %considered to characterize 
 regarded as
 the degree 
 of the overlap
 of the 
 volumes
  of the 
two colliding 
 nuclei, 
 %determined 
 characterized
 by the impact parameter.
 % \ct{GlauberMCrev}. 
 The most central collisions 
correspond, therefore, to the lowest centrality while the larger
 centrality 
 %define 
  to the
 more peripheral collisions. The centrality is closely 
related to the number of nucleon participants determined using Monte 
Carlo Glauber calculations.
 Hence,
 % so that 
 the largest number of participants 
contribute to the most central heavy-ion collisions. 
 %Hence t
 The centrality 
 is 
 thus
 related to the
   amount of 
 %%the
  energy released in the collisions, \ie\
   to
 the effective energy, $\eNN$.
 %, which, 
 The latter,
in the framework of the 
 proposed approach,
 can be 
defined as 
a fraction of the c.m. energy available in a collision 
according to the centrality, $\alpha$: 

\begin{equation}
\eNN = \sqrt{s_{NN}}(1 - \alpha).
\label{Eeff}
\end{equation}
 Conventionally, the data are divided into 
 %classes of centrality, or 
 centrality intervals, so that $\alpha$ is the average centrality 
 per 
 %for 
 %the 
centrality interval, 
\eg\  $\alpha = 0.25$ 
for the 
centrality interval of 20\%--30\% centrality.

Then, for 
 the 
effective
c.m.  energy $\eNN$, 
   Eq.~(\ref{pmult})  reads
 
\begin{eqnarray}
 \nonumber
 & & \frac{2\, N_{\rm ch}}{N_{\rm part}} =
  N^{pp}_{\rm ch} \,
   \frac{\rho(0)}{\rho_{pp}(0)} \,
  \sqrt{1-\frac{2 \ln 3}{\ln\, (4.5\, \eNN/m_ p)}}\,,  \\
& & \eNN=\spp/3\,,
\label{pmultc}
\end{eqnarray} 
 where
$\rho(0)$ 
is the  midrapidity density
 in central nucleus-nucleus collisions measured 
at  $\sNN=\eNN$.

 In fact, each of the scalings described by Eqs.(\ref{pmult}) 
 and 
 (\ref{Eeff}) regulates a
particular physics ingredient used in the modelling of the PDE approach.
The scaling introduced by Eq.(\ref{pmult}) embeds the constituent 
quark 
model,
which leads to establishing a similarity between hadronic and 
nuclear collisions. 
 The scaling driven by Eq.(\ref{Eeff}) 
addresses the
energy budget effectively retained in the most central collisions while 
considering the global variables from 
noncentral collisions.
 %% appealed to define the energy budget
 %% effectively retained for multiparticle production in the 
 %% most central 
 %% collisions to determine the variables obtained from 
 %%centrality data. 
\\

%Fig.1
\begin{figure*}%\sidecaption
\begin{center}
\resizebox{0.6\textwidth}{!}{%
  \includegraphics{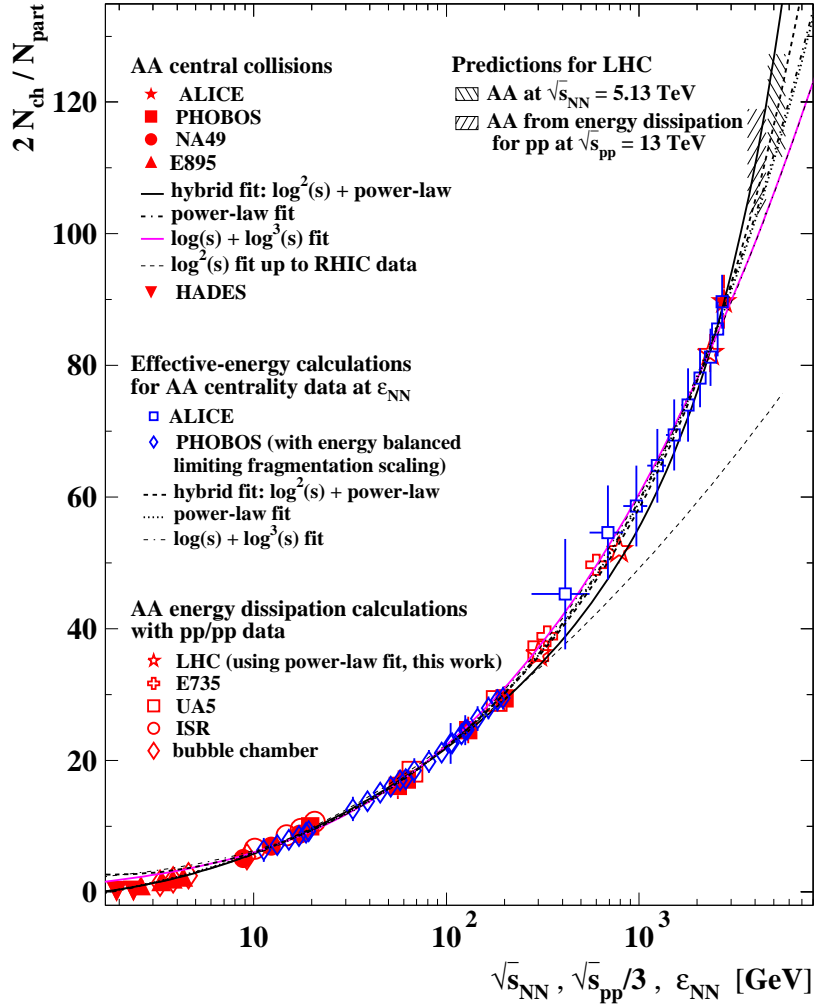}
}
\end{center}
\caption{\small 
The energy dependence of the 
 charged particle mean multiplicity per participant pair.
 The large solid symbols show the measurements from the most central 
nucleus-nucleus (AA) collisions 
 given 
 as a function
 of the nucleon-nucleon c.m. energy, $\sNN$.  
 The calculations 
 by Eq. (\ref{pmult})
 based on 
$pp/\pbp$ data at the c.m. energy $\spp=3\,\sNN$ 
 are shown \vrs\ $\spp/3$  by large open symbols. 
 The small open symbols 
show the  
AA data at 
different 
centralities 
 %by using Eq. (\ref{pmultc}) at 
 as a function the effective energy 
 $\varepsilon_{NN}$ (Eq.~(\ref{Eeff})).
 The RHIC centrality data are shown 
 after 
 removing   
 the energy-balanced 
limiting fragmentation scaling ingredient, 
 %from the calculations of Eq. (\ref{pmultc}), 
 while 
 %the calculations do not take into account 
 this 
 ingredient 
  is not taken into account 
 for the 
 LHC centrality data (see text). 
 The multiplicity data of the
  most-central AA collisions 
 are measured by 
 the ALICE experiment at LHC \ct{alice-mult},
 by the PHOBOS experiment at RHIC 
 \ct{ph-rev,phobos-all},
  by the NA49 experiment at CERN SPS \ct{na49-mult}
 and by the E895 experiment at AGS \ct{agsmult}
 (for the latter see also \ct{ph-rev}). The low-energy HADES measurements 
at GSI
are
taken from
\ct{hades}.
 The centrality data are taken from the measurements by the PHOBOS 
experiment 
at 
RHIC 
\ct{phobos-all} and by the ALICE experiment at the LHC 
\ct{alice-mult,alice-mult2}.
 The values obtained from
Eq.~(\ref{pmult}) for the AA mean multiplicity are based on:
 nonsingle diffractive  $\pbp$ collisions at FNAL by the E735 
experiment
\ct{pp-mult-rev,e735-conf-mult},
 at CERN by
 the UA5 experiment
 at $\spp=$ 546 GeV \ct{ua5-546}
 and $\spp=$ 200 and 900 GeV \ct{ua5-zp}; 
 {\it pp} collisions
 from CERN-ISR
   \ct{isr-nsd},
 and from the inelastic data
 from the bubble chamber experiments
 \ct{fnalmult,bubblechamber-ammosov,bubblechamber-bromberg},
the latter having been compiled and analyzed
 in
 \ct{eddi}.
 The LHC 
 multiplicities 
 in $pp$ interactions 
 are  calculated using the
 hybrid 
 fit
 obtained here, 
 Eq.~(\ref{ppexp}).
 The solid and the dashed-dotted 
show, 
correspondingly,  the hybrid 
fit,
$ -0.577 + 0.394 \ln(\sNNq) 
  + 0.213 \ln^2(\sNNq)+0.005 \sNNq^{0.551}$, 
and   the power-law fit,
  $-6.72+5.42\,s_{NN}^{0.18}$,
 to the most central AA 
data.
 The thin dashed
 line shows the second-order log fit
 $-0.35+0.24\ln(s_{NN})+0.24\ln^2(s_{NN})$  to the most central AA data up 
to the top 
RHIC energy \ct{edward,edwarda}.
  The dashed and the dotted lines  show, correspondingly,  the 
hybrid 
fit, 
$3.04-1.4\ln(\eNN) +1.12\ln^2(\eNN)+0.032\, \eNN^{0.848}$, and the 
power-law fit,  $-6.62+5.43\,\eNN^{0.362}$, to the centrality AA 
 data. 
 The 
 % grey 
  pink 
 solid line 
 and the thin dashed-dotted line show the fits
  $0.72+0.75\ln(\sNNq) +0.019\ln^3(\sNNq)$
 and $1.7+2.36\ln(\eNN)+0.152\ln^3(\eNN)$ 
  to the most central  collision and centrality AA data, respectively
(see text).
  The right-inclined hatched area  
  shows the
prediction for 
heavy-ion 
collisions at $\sNN=$~5.13 TeV and the left-inclined hatched area 
  gives the 
prediction 
 expected from 
 $pp$ collisions at $\spp=$~13 TeV.
 }
\label{fig:nvss}       % Give a unique label
\end{figure*}

 %Fig. 2. pp 
 \begin{figure*}%\sidecaption
\begin{center}
 \resizebox{0.6\textwidth}{!}{%
 \vspace*{.2cm}
\includegraphics{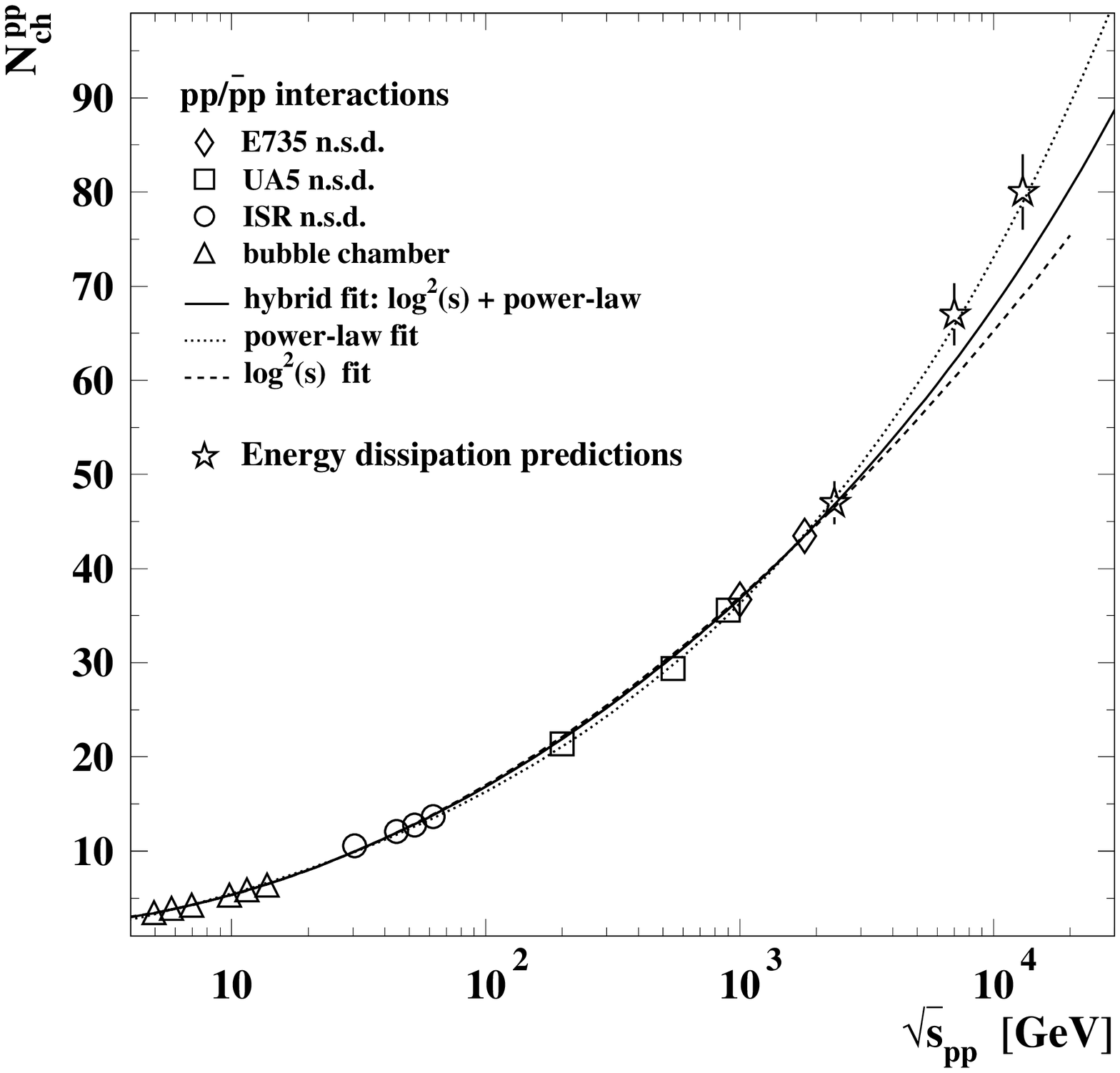}
}
\end{center}
\caption{\small
The c.m. energy dependence of the 
 charged particle
 mean multiplicity 
in $pp/\pbp$ collisions. The measurements are taken from:
 nonsingle diffractive  $\pbp$ collisions at FNAL by the E735 experiment
\ct{pp-mult-rev,e735-conf-mult},
 at CERN by
 the UA5 experiment
 at $\spp=$ 546 GeV \ct{ua5-546}
 and $\spp=$ 200 and 900 GeV \ct{ua5-zp}; 
 {\it pp} collisions
 from CERN-ISR
   \ct{isr-nsd},
 and from the inelastic data
 from the bubble chamber experiments
\ct{fnalmult,bubblechamber-ammosov,bubblechamber-bromberg,eddi,bubblechamber-benecke}.
  The solid line 
shows  the hybrid 
fit,
$1.60-0.03\,
 \ln(\sppq) 
  + 0.18\, \ln^2(\sppq)+0.03\, \sppq^{0.29}$,
 the dotted line shows 
 the power-law fit, 
  $-7.36+6.97\,s_{pp}^{0.133}$,
 and 
 the dashed
 line shows the second-order log fit,
 $3.18-0.57\ln(s_{pp})+0.216\ln^2(s_{pp})$.
The open stars show
 the predictions by the PDE 
 % participant dissipating
 % energy 
 approach with the error bars of 5\% uncertainty.
}
\label{fig:npp}       % Give a unique label
\end{figure*}

 %\nopar
 \section{Multiplicity c.m. energy dependence in central nuclear 
 and $pp/\pbp$ collisions}
 \label{sec:nvss}
 %{\bf 3. } 
 Figure \ref{fig:nvss} shows the c.m. energy dependence of the 
multiplicity 
measured in head-on nucleus-nucleus collisions (solid symbols) in the 
energy range 
 of
$\sNN = 2$~GeV to 2.76~TeV.
 Given 
 the fact that 
the measurements support
the second-order logarithmic 
dependence 
on $\sNN$  up to the top RHIC energy \ct{edward,phobos-all} 
while the power-law dependence is obtained for the LHC data 
\ct{alice-mult},
we fit the head-on data by the ``hybrid'' fit function, 

\begin{eqnarray}
\nonumber
 \frac{2\, N_{\rm ch}}{N_{\rm part}} 
 =  (-0.577\pm 0.177)  + (0.394\pm 0.094)\,
 \ln(\sNNq/s_0) & \\ 
 \nonumber
  + (0.213\pm 0.014)\, \ln^2(\sNNq/s_0) &\\
+(0.005\pm 0.009) 
\,  (\sNNq/s_0)^{(0.55\pm 0.11)}. &
\label{hybnaa} 
\end{eqnarray}
 %
 % This is
 % similar to the successful hybrid fits considered by us 
 % earlier for the  charged particle 
 % density and the 
 % transverse energy 
 % density at midrapidity \ct{edward2}.
 Here, $s_0=1$~GeV$^2$. In the following, the factor $s_0$ is taken the 
same 
in all fit functions and omitted for brevity.
This fit is shown in 
Fig.~\ref{fig:nvss}
by the solid line. 
 Note that from 
 the theoretical description point of view, 
 the logarithmic dependence is considered to characterize 
 the 
 fragmentation 
 source(s) 
 while the power-law behavior is believed to come from the 
 gluon-gluon 
 interactions~\ct{wolschin}; for a review, see \ct{AHPreview}.

We also fit the head-on  collision multiplicities with the power-law 
function. 
 The power-law dependence of the multiplicity is  expected in 
different theoretical approaches \ct{levin,capella,percol1} and the data 
from 
nuclear and 
$pp/\pbp$ collisions  seem to follow this 
type of
behavior at higher energies \ct{pp-mult-rev,alice-mult}.
 The
 power-law fit gives:
\begin{equation}
%\nonumber
 \frac{2\, N_{\rm ch}}{N_{\rm part}} 
=  (-6.72\pm 1.44)+(5.42\pm 
1.11) \,  \sNNq^{(0.18\pm 0.02)}, 
\label{expnaa} 
\end{equation}
 and is shown by the dashed-dotted line in 
Fig.~\ref{fig:nvss}.
 
Recently, it was shown \ct{wolschin-ln3} that the multiplicity of the 
gluon-gluon 
interactions are better 
described within a nonequilibrium statistical relativistic diffusion 
model using log$^3(\sNNq)$ dependence.  Using this behavior we fit the 
c.m. energy dependence by the corresponding fit function, 
 %
%\begin{equation}
\begin{eqnarray}
\nonumber
 \frac{2\, N_{\rm ch}}{N_{\rm part}} 
=  (0.72\pm 1.85)  + (0.75\pm 0.39)\, \ln(\sNNq) & \\ +(0.019\pm 
0.002)\,\ln^3(\sNNq). &
\label{ln3} 
 %\end{equation}
\end{eqnarray}
 Here, the linear-log term reflects the multiplicity from the 
fragmentation sources, as noticed above. The fit is shown by the 
 %grey
 pink 
 solid 
line in Fig.~\ref{fig:nvss}. The fit is made starting the lowest 
NA49 energy 
in 
order to match the LHC multiplicity. One can see that the fit seems to be
indistinguishable from the power-law function for the entire fit range, 
and is 
slightly below the power-law behavior above the current LHC data. 
 Some enhancement in the low-energy range is  expected due to no gluonic
source considered to be present at these energies.

 In addition to these  fits, 
 we show the 
 $\log^2(\sNNq)$-fit  \ct{edward,edwarda} 
up to the top 
RHIC energy  
 (thin dashed 
 line). 
 One can see that the power-law fit well describes the data and 
is almost indistinguishable from the hybrid fit up to the LHC data. 
Some 
minor deviation between the two fits can be seen in the  range 
from the top RHIC energy to the LHC energy. 
 %Meanwhile,
   Meantime,
 the second-order log polynomial lies below the data for $\sNN>200$~GeV.
 This observation supports a possible transition to  a new regime in
heavy-ion
collisions at $\sNN$ 
  of
 about 1~TeV, as indicated earlier 
in 
the studies of 
pseudorapidity particle  and transverse 
energy  densities at midrapidity \ct{edward2}.

  Addressing now Eq.~(\ref{pmult}), we calculate the 
 mean multiplicity $N_{\rm {ch}}/(N_{\rm part}/2)$ for nucleus-nucleus 
 interactions 
 using
 the $pp/\pbp$ measurements.
  The calculated values are shown in 
Fig.~\ref{fig:nvss} 
by large open 
symbols. 
 The 
 rapidity 
 density $\rho_{pp}(0)$ and the multiplicity $N_{\rm{ch}}^{pp}$ 
 are taken from the existing data \ct{pdg14} or, where not available, 
calculated 
using the  
 corresponding experimental $\spp$ fits\footnote{The 
 powewr-law
 fit, Eq.~(\ref{ppexp}), 
 is used for  $N_{\rm{ch}}^{pp}$, while $\rho_{pp}(0)$ is calculated using
the 
linear-log fit 
$\rho_{pp}(0)=-0.308+0.276\, \ln(s_{pp})$ \ct{pp-mult-rev} and the 
 power-law fit by CMS \ct{cms276c},  $\rho_{pp}(0)=-0.402+s_{pp}^{0.101}$,
 at $\sqrt{s_{pp}}\leq$~53~GeV and at $\sqrt{s_{pp}}>$~53~GeV, 
respectively.}
at $\spp=3\, \sNN$, 
  in  accordance with the 
 approach considered here.
The $\rho(0)$ values are as well 
taken from the 
measurements in central heavy-ion collisions wherever available, while for 
the nonexisting data the experimental fit\footnote{The  
 linear-log fit  $\rho(0) = -0.33+0.38\ln(\sNNq)$ \ct{edward,phobos-all} 
is applied at $\sNN\leq 63$~GeV, and the  power-law 
fit $\rho(0) = 
0.73\, \sNNq^{0.155}$ is applied above 63~GeV  as recently reported by 
ALICE using the measurements up to 
$\sNN=$~5.02 TeV \ct{alice5020}.}
 is used.

One can see that the calculated $N_{\rm ch}/(0.5 N_{\rm
part})$ values follow
 the  measurements from  nucleus-nucleus collisions 
 at $\sNN$ from a few 
GeV up to the TeV LHC 
 energy. 
  %Slight deviation seen in the calculations using the LHC 
 %$pp$ data 
 %above $\sNN=$~600~GeV, \ie\ 
 %for $\spp > $~1.8 TeV, 
 %can be explained by no data 
 %on $N_{\rm ch}^{pp}$ being available above the Tevatron energy 
 %of  $\spp=$~1.8~TeV 
 %and the use of the
 %fit of Eq.~(\ref{pphybr})
 %instead (see below).
 The observed agreement between the heavy-ion measurements of $N_{\rm 
ch}/(N_{\rm part}/2)$ and the values obtained from the $pp$-based 
calculations points to the 
universality of the multiparticle production process in  different 
types of collisions.

 Solving Eq.~(\ref{pmult}) for 
the 
mean 
multiplicity  
  $N_{\rm
ch}^{pp}$  
 in $pp$ collisions,
 we estimate its values 
 for $\spp>$ ~2~TeV to be about
 47 at $\spp=$~2.36 TeV, 67 at 7 TeV, and 79 at 
13
TeV with 
5\% 
 uncertainties. 
  Here for the calculations, one uses the fit to the heavy-ion midrapidity 
density 
data $\rho(0)$, 
 as described above,
and the fit by ALICE to 
 %the average of the above-obtained fits, 
 % Eqs.~(\ref{hybnaa}) 
 % and 
 %((\ref{expnaa}), to 
 the head-on heavy-ion data on 
the mean multiplicity \ct{alice-mult} [similar to the 
results for the fits of  Eqs.~(\ref{hybnaa}) 
  and 
 (\ref{expnaa})], 
   along with 
the LHC  
 %non-single diffractive 
measurements 
 \ct{cms900236,alice900236,cmspp7pTn,atl9002367}  
 of the 
pseudorapidity density 
 $\rho_{pp}(0)$. 
 %at midrapidity 
 %
 The 
 calculated
 values of $N_{\rm ch}^{pp}$ 
 are shown as
a function of
$\spp$
 by open stars in Fig.~\ref{fig:npp}, 
 along with the existing multiplicity measurements from 
 $pp/\pbp$ collisions.

 The 
measured  $N_{\rm ch}^{pp}$ dependence on $\spp$ 
in the energy range 
spanning the interval between 
 a  
few GeV to 
1.8 TeV are fitted with the   
 power-law, second-order log-polynomial and the hybrid functions.
 % Similar to the above hybrid fit to the head-on nucleus-nucleus 
 % data, 
 % Eq.~(\ref{hybnaa}), t
 The hybrid and the power-law fits read,

 \begin{eqnarray}
\nonumber
 N_{\rm ch}^{pp} 
= (1.60\pm 0.23)+ (-0.03\pm 0.10)\,
 \ln(\sppq) & \\ 
  + (0.18\pm 0.01)\, \ln^2(\sppq)+(0.03\pm 
0.02) \,  \sppq^{(0.29\pm 0.06)}, &
\label{pphybr} 
\end{eqnarray}
 and 
 \begin{equation}
 %\nonumber
 N_{\rm ch}^{pp} 
=  (-7.36\pm 0.16)+ (6.97\pm 0.12)\,  \sppq^{(0.133\pm 0.001)}, 
\label{ppexp} 
\end{equation}
 correspondingly.
  % is 
 % used to fit  $pp/\pbp$ data by combining the 
 % log$^2(s)$-polynomial and the 
 % power-law fit functions.

 From Fig.~\ref{fig:npp} one can conclude that the available  data do 
not give any preference to one or another fit 
function. 
  This is similar to the pre-LHC observations 
where the power-law fit   were found to be indistinguishable from 
 the log$^2$ polynomial fit
 at
$\spp>53$~GeV \ct{pp-mult-rev}. 
 Interestingly,  these two functions are also 
found to fit equally well the 
nonsingle 
diffractive 
midrapidity density, as obtained by CMS: cf. fits in \ct{cmspp7pTn} {\vrs} 
those in \ct{cms276c}.    
 The fit functions 
 start to  deviate 
 from each other
at the c.m. energy above a few TeV but still not far one from another 
even at $\spp \sim 10$~TeV. 
 % The log$^2(s)$-polynomial fit is  from the power-law fit even 
  % for   
  % $\spp>2$~TeV.
 This   
may point to apparently no change in the 
multihadron production in $pp$ interactions up to the 
 %top  
 highest
 LHC 
 energy, in contrast to 
a new regime possibly 
 %occurred 
 emerging 
 at $\sNN\approx 1$~TeV in 
heavy-ion collisions. 

 It is remarkable how well  the 
PDE  predictions 
on $N_{\rm ch}^{pp}$ 
at $\spp >2$~TeV 
 % based on the PDE
 %energy-dissipation 
  % approach 
 %of quark particuipants
 follow the  
power-law fit made to the measurements  at $\spp \leq 1.8$ 
TeV.
  This and the above-indicated ``no change'' in the hadroproduction in 
$pp$ 
collisions as soon as one 
moves to  
 TeV energies
 %%, both 
 are in an agreement with the 
prediction 
 \ct{edward} 
which 
seems 
 to be the only successful 
  %description 
 one 
 %made 
 for 
 %of 
 the midrapidity 
density in $pp$ collisions at $\spp=7~$TeV \ct{CMSwe}.
   \\

%Fig.3
%
 \begin{figure*}%\sidecaption
\begin{center} 
%\resizebox{0.5\textwidth}{!}{%
 \resizebox{0.7\textwidth}{!}{%
  \includegraphics{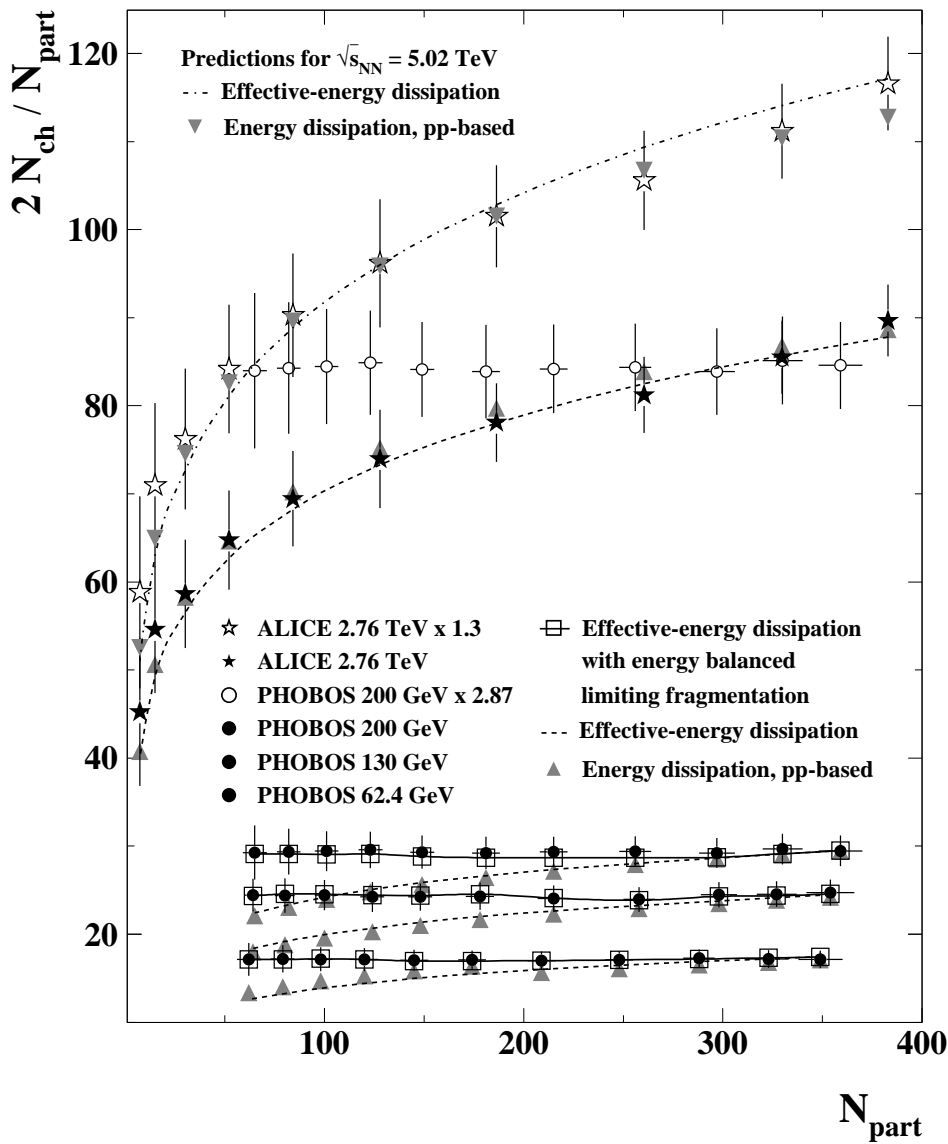}
}
\end{center}
\caption{\small
 The 
 charged particle 
 mean multiplicity 
per participant 
pair as a function of the number of participants, $N_{\rm part}$.
 The 
 solid 
 circles show the dependence 
 measured in
 AuAu collisions at RHIC  by 
the PHOBOS experiment at $\sNN=62.4, 130$ and 200~GeV  
\ct{phobos-all} (bottom to top). The 
 solid
 stars show the measurements 
 from 
PbPb collisions at the LHC by the ALICE experiment at $\sNN=2.76$~TeV  
\ct{alice-mult2}.
 The 
 triangles show the calculations
by Eq.~(\ref{pmultc}) using $pp/\pbp$ data.
 The  dashed  lines represent  
the
 calculations
 within the 
 effective-energy 
 approach 
 %calculations    
 based on the hybrid fit, Eq. (\ref{hybnaa}),
 to the c.m. energy
dependence of the 
 mean multiplicity in the most central heavy-ion collisions shown in 
Fig.~\ref{fig:nvss}. The dashed-dotted line show the predictions for 
$\sNN=5.02$~TeV using the average of the fits,  Eq. (\ref{hybnaa}) and  
Eq. 
(\ref{expnaa}).
 The 
 open
 squares show
the effective-energy 
  calculations
 which include 
 the energy-balanced
limiting fragmentation scaling
 (see 
text); 
the solid lines connect the calculations
 to guide the eye.  
 The 
 open 
 circles show the PHOBOS measurements at $\sNN=200$~GeV 
multiplied by 
2.87.
 The 
 open 
 stars show the ALICE measurements at $\sNN=2.76$~TeV 
multiplied by 
1.3.
 }
\label{fig:nvsnpart}       % Give a unique label
\end{figure*}

 %\nopar
 \section{Multiplicity centrality dependence} 
 \label{sec:nvsnpart}
 %{\bf 4. } 
 In this section, we address  
  the point whether
the 
centrality dependence of the
 mean multiplicity 
from heavy-ion 
experiments 
 %to be 
 is
 described by 
Eq.~(\ref{pmultc}), similarly to the midrapidity pseudorapidity density 
 in \ct{edward2}.
In Fig.~\ref{fig:nvsnpart}, we show the $N_{\rm part}$-dependence of  
$N_{\rm ch}/(N_{\rm 
part}/2)$. The data are taken from the measurements by
 the
 PHOBOS 
experiment at RHIC 
\ct{phobos-all} 
and by 
 the
 ALICE experiment at LHC \ct{alice-mult2}.
 %The PHOBOS data at  $\sNN=200$~GeV
 %multiplied by 2.87  
 %are also shown to allow comparison with the LHC data and the
 %current calculations. 
 The solid triangles show  
 the 
 estimations using 
 Eq. (\ref{pmultc}). 
 As above,    
 in the case of  
 the $\sNN$-dependence,
 the
 rapidity
 densities $\rho_{pp}(0)$ and  $\rho(0)$, and the multiplicity 
$N_{\rm{ch}}^{pp}$ 
 are taken from the existing data \ct{pdg14} or, where not available, are 
 calculated from the 
 %above-noticed fits
fits
 described above. 
 According to the
 consideration 
 developed here,  
  $\rho(0)$ is taken at $\sNN=\eNN$, and 
  $\rho_{pp}(0)$ and $N_{\rm{ch}}^{pp}$ are 
taken at $\spp=3\, \eNN$.

 One can see that 
 the calculations,
  which  
 are  
 driven by the centrality-defined effective c.m. energy $\eNN$,   
 well reproduce 
 the 
 LHC data
 except slightly underestimating 
 a couple of the most peripheral 
  measurements. 
 For the RHIC data, however, the  
 difference between the calculations and the measurements 
 is 
 visible 
 already 
 for
 medium centralities,  
 \ie\ for 
 more central collisions.
  These observations are also interrelated with  the difference observed 
in the measurements at RHIC \vrs\  
those from 
LHC. Indeed, at RHIC, the participant-pair-normalised mean 
multiplicity is found to be 
independent   of centrality, 
 while 
 a
 decrease with centrality, or monotonic increase with $N_{\rm part}$, 
is observed at the  LHC. 
 %The 200 GeV RHIC data well clarifies on this being multiplied by 
 %a factor to be compared with the LHC data. 
 This becomes even clearer when the  200 GeV PHOBOS data
 are multiplied by a factor of 2.87 
 (open circles in Fig.~\ref{fig:nvsnpart})
which allows matching the ALICE data 
from the highly central collisions.

 In  Fig.~\ref{fig:nvsnpart},  
the 
above-obtained c.m. energy fit,  Eq.~(\ref{hybnaa}), 
made 
to the {\it head-on} collision data, 
is applied 
 to the {\it centrality} measurements at $\sNN=\eNN$ and the results are 
shown by the 
dashed lines.\footnote{Similar results are obtained from 
Eq.~(\ref{expnaa}).}  
 The observations 
 made for the  calculations   
 are valid 
 here as well.
 This 
 points to 
the complementarity of central collisions and centrality data 
  once
 the calculations are made in the c.m. effective-energy $\eNN$ 
 terms.
 %alike
  % similarly 
  % to
  % the 
  % observations 
  % made  for the measurements of the 
  % pseudorapidity and 
 % transverse-energy densities at midrapidity 
 % as discussed earlier.
 % \ct{edward2}.

To clarify the observed differences, 
 in the 
 following
 sections 
the 
distributions 
of the 
pseudorapidity density 
are investigated in the 
context of the PDE picture considered here.  
\\

%Fig.4
\begin{figure*}%[h!]
\begin{minipage}[h]{0.48\textwidth}
\center{
\includegraphics[width=1.1\linewidth, height=0.35\textheight]{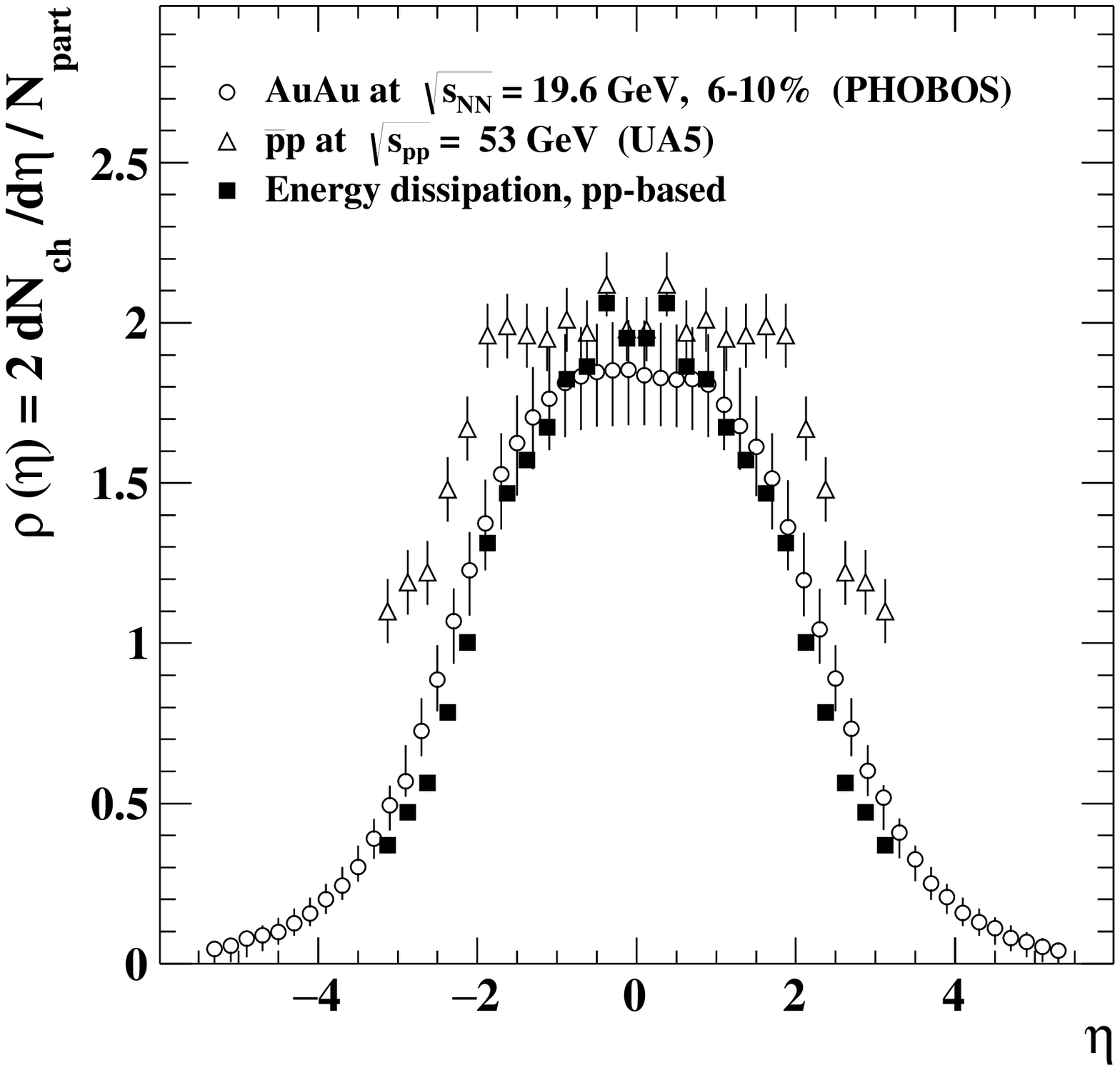}
}
a) \\
\end{minipage}
%\end{center}
\hfill
\begin{minipage}[h]{0.48\textwidth}
\center{
\includegraphics[width=1.1\linewidth, height=0.35\textheight]{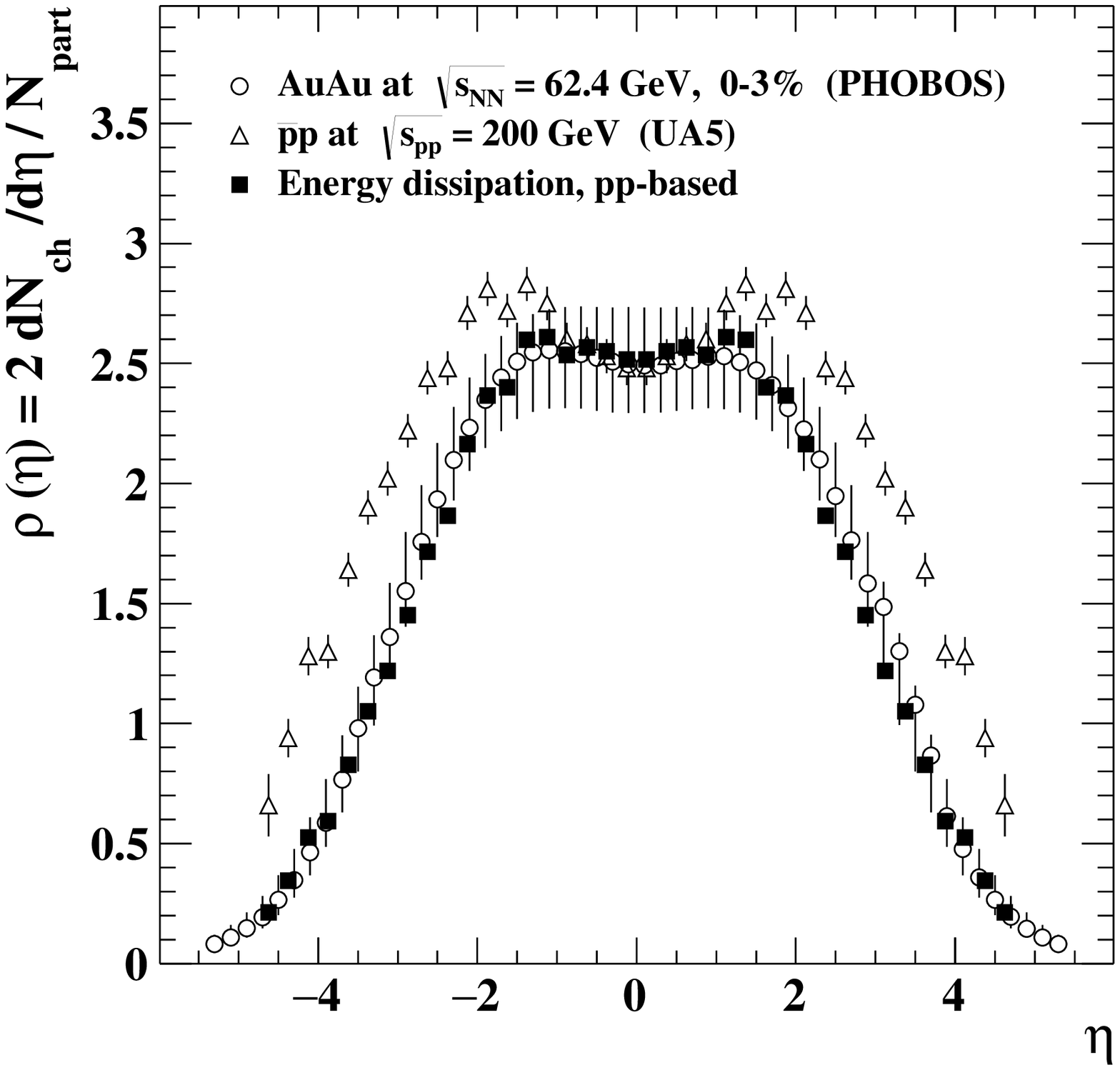}
}
b) \\
\end{minipage}
\hfill
\begin{minipage}[h]{0.48\textwidth}
\center{
\includegraphics[width=1.1\linewidth, height=0.35\textheight]{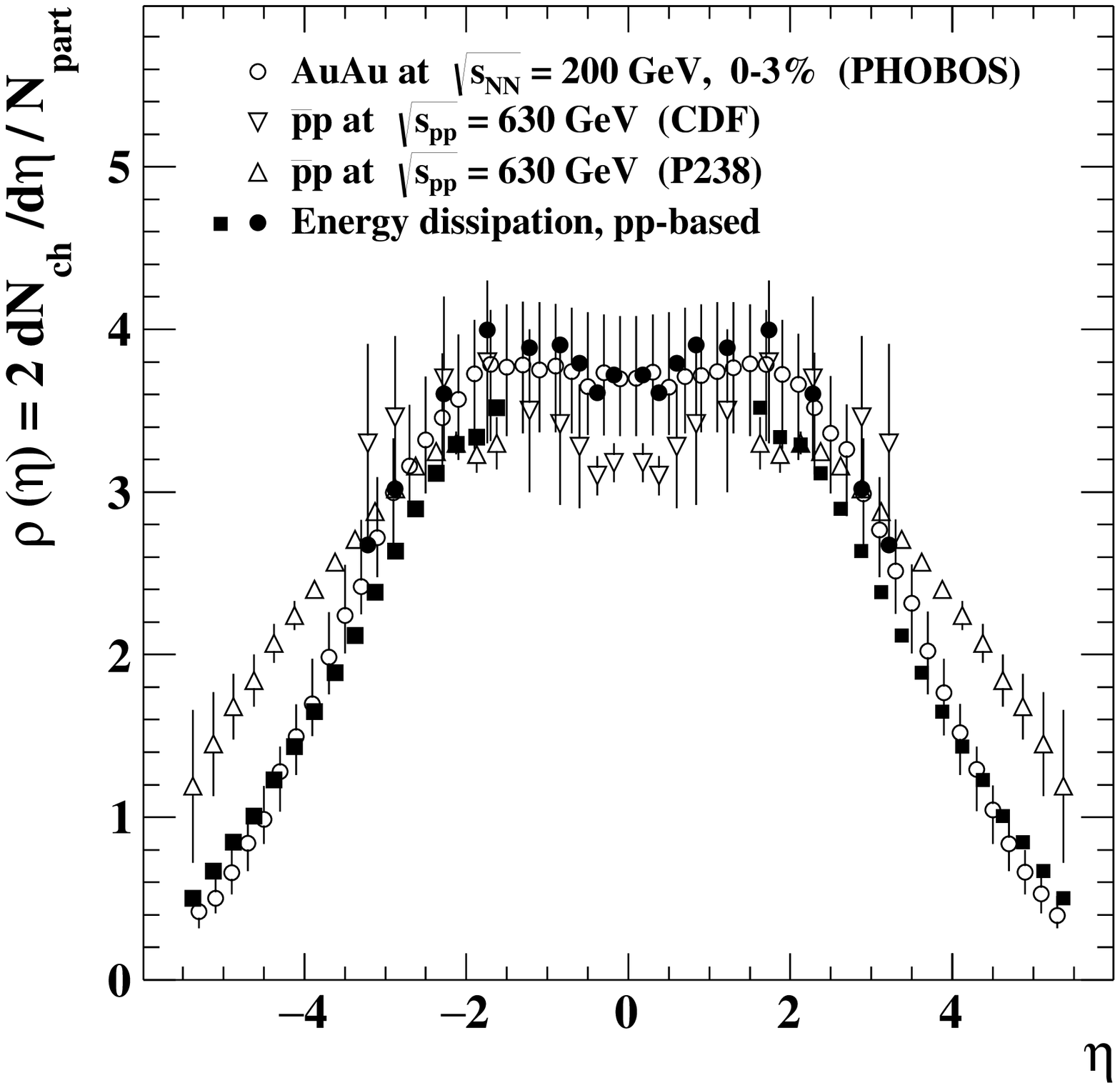}
}
 c) 
\\
\end{minipage}
\hfill
\begin{minipage}[h]{0.48\textwidth}
\center{
\includegraphics[width=1.1\linewidth, height=0.35\textheight]{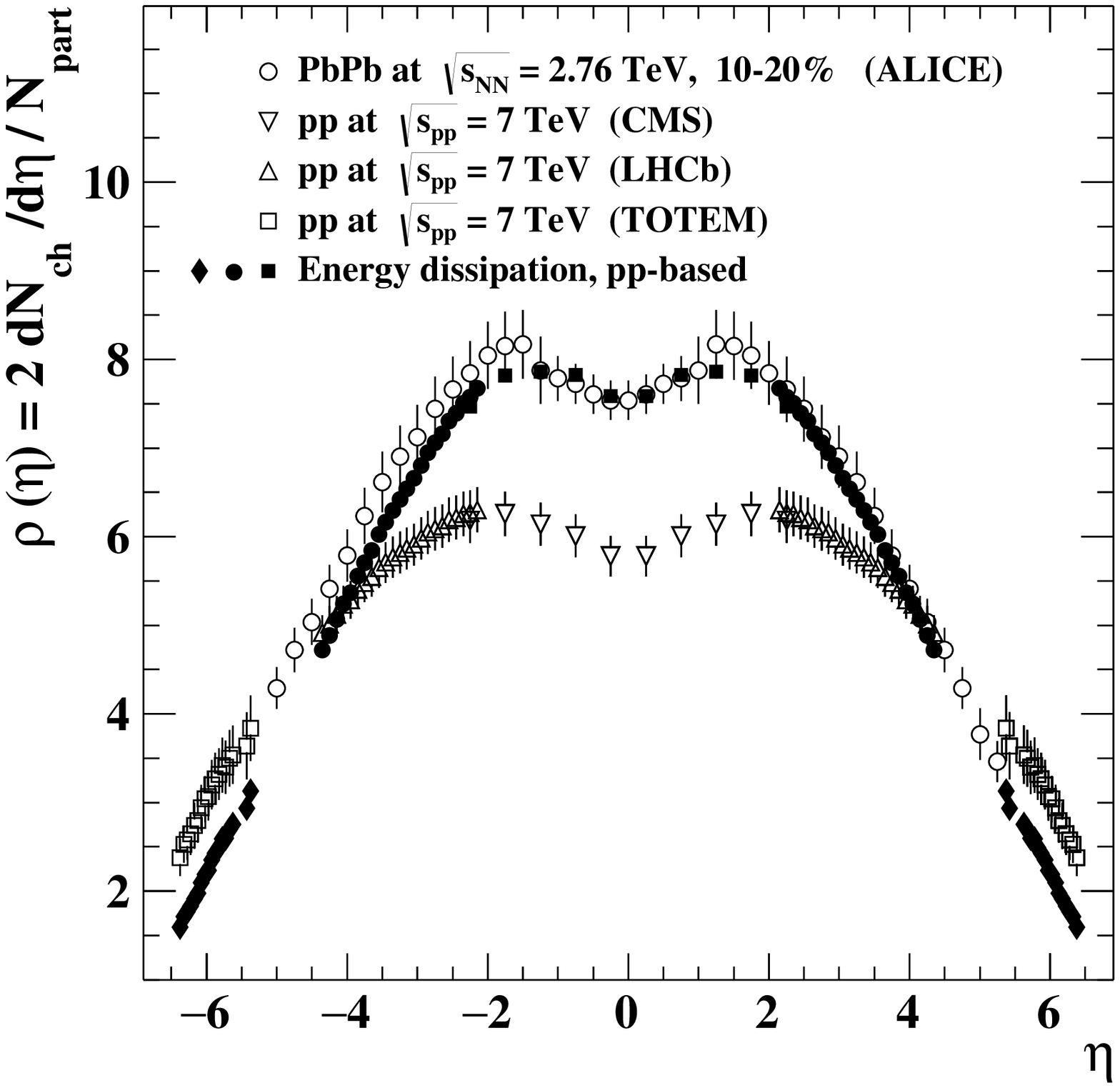}
}
 d) 
\\
\end{minipage}
\caption{\small
 The 
 pseudorapidity  distributions of 
charged particle pseudorapidity 
density  per 
 participant pair.
 The 
 open 
 circles show the 
measurements by
the PHOBOS experiment in
AuAu collisions at RHIC at (a) $\sNN=19.6$~GeV, (b) 62.4~GeV and (c) 
200~GeV
\ct{phobos-all}, and 
(d) 
by the ALICE experiment in PbPb collisions at the LHC at $\sNN=2.76$~TeV 
\ct{alice-mult}. The 
 open 
 triangles  show the distributions 
measured
in $\pbp$ interactions
by the UA5 experiment at $\spp=53$~GeV at the ISR and at  $\spp=200$~GeV 
at 
the SPS 
\ct{ua5-zp-rap},  
by the P238 experiment at the SPS \ct{p238} and by the CDF experiment at 
the 
Tevatron 
\ct{cdf630} 
at  $\spp=630$~GeV, and in $pp$ interactions by the CMS \ct{cmspp7pTn}, 
LHCb 
\ct{lhcb7rap} 
and TOTEM 
\ct{totem7} experiments 
at $\spp= 7$~TeV at the LHC.
 The 
 solid markers show the calculations by 
Eq.~(\ref{rapdist}) using $pp/\pbp$  data at $\spp \approx  3\,\sNN$ or 
$3\,\eNN$. 
 Apart from 
 the CMS data, 
 the negative-$\eta$ data points for $pp/\pbp$ 
interactions are the 
reflections of the measurements taken in the positive-$\eta$ region.   
\label{fig:rapmcentr}}
\end{figure*}

%Fig.5
\begin{figure*}[t!]
\begin{minipage}[h]{0.48\textwidth}
\center{
 \includegraphics[width=1.1\linewidth, height=0.35\textheight]{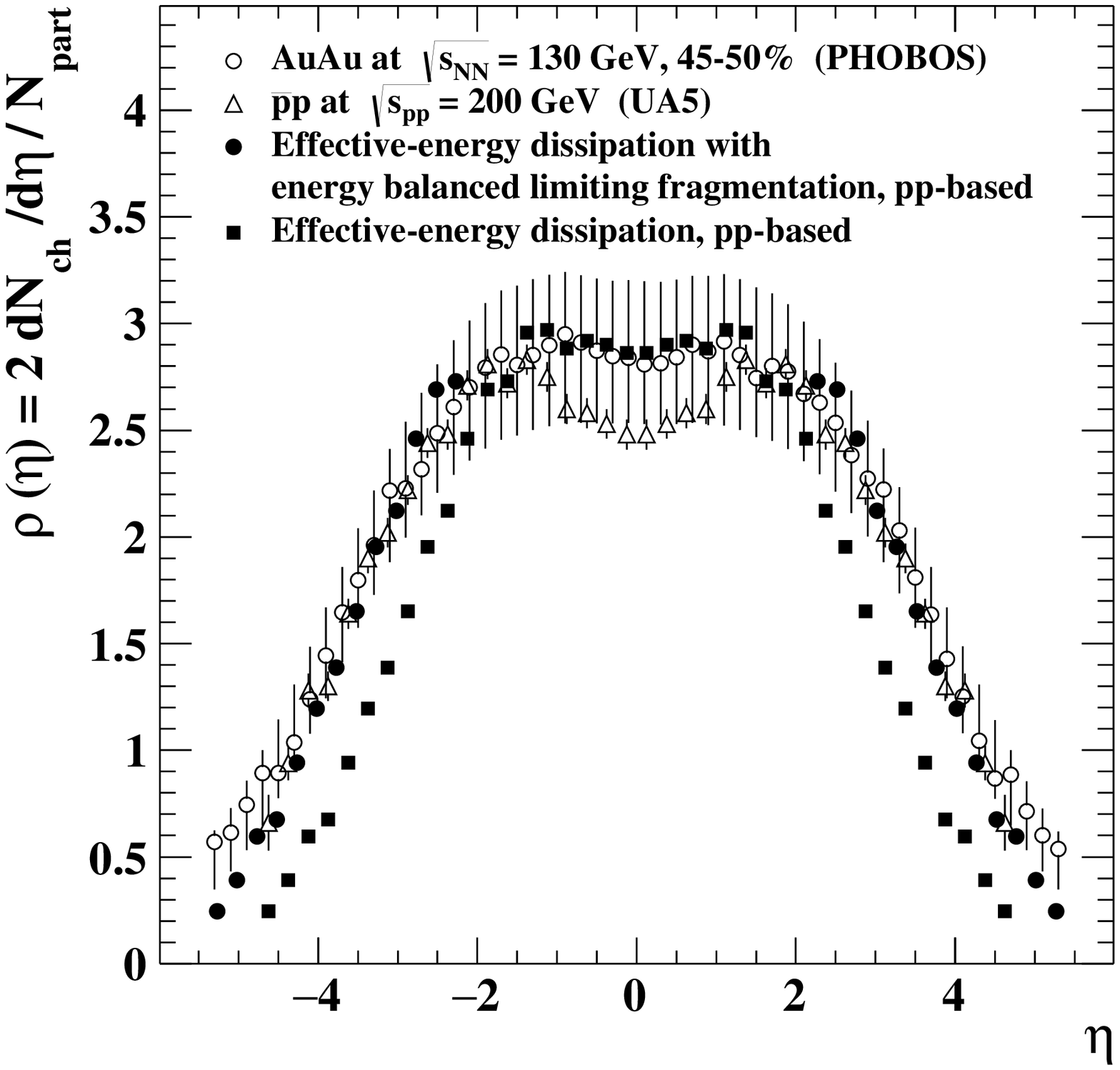}
}
a) \\
\end{minipage}
\hfill
\begin{minipage}[h]{0.48\textwidth}
\center{
 \includegraphics[width=1.1\linewidth, height=0.35\textheight]{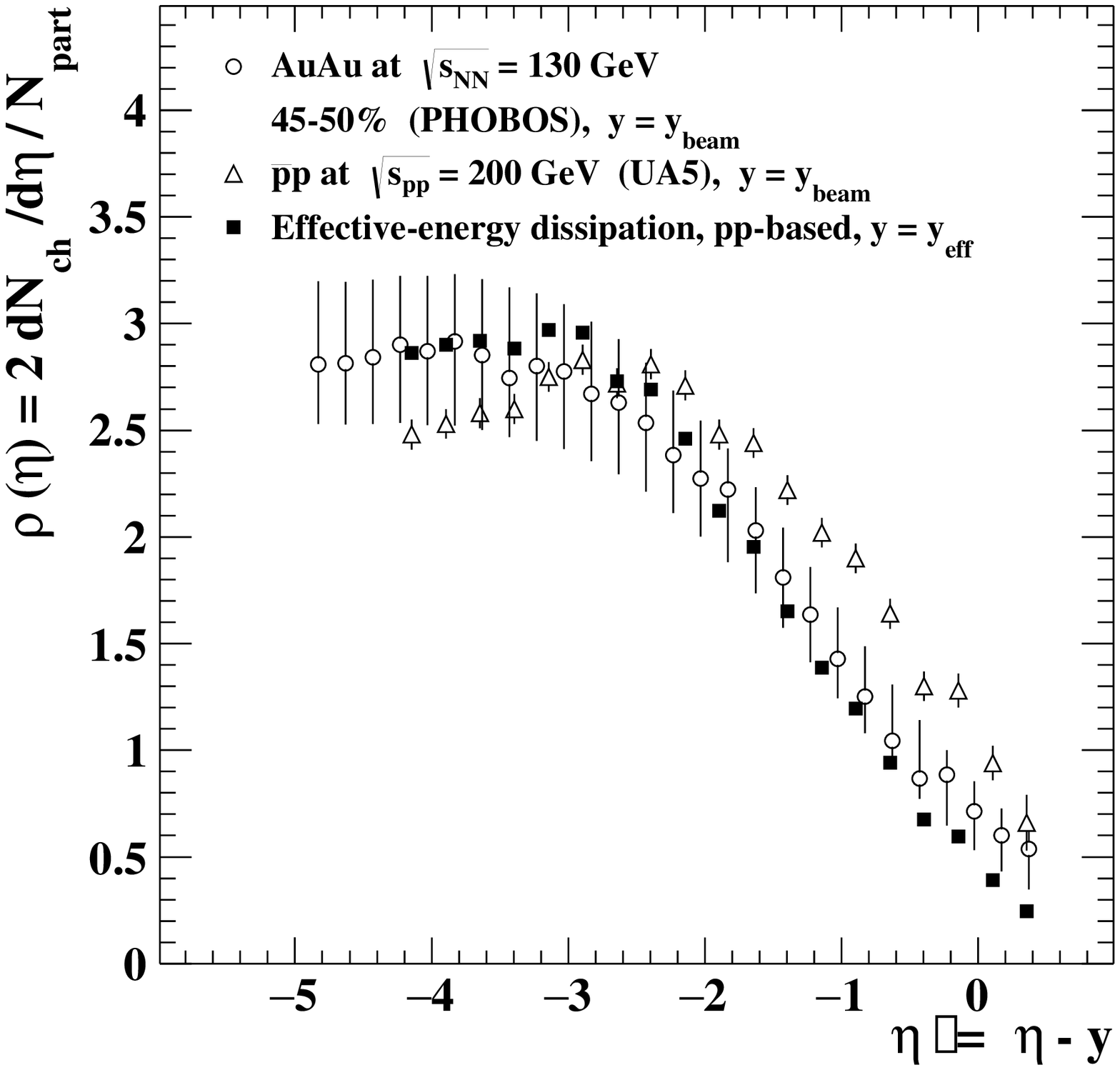}
}
b) \\
\end{minipage}
\caption{\small
 (a) The 
 charged particle pseudorapidity density 
per participant pair
as a function of pseudorapidity.
 The 
 open 
 circles
show the distribution
measured in
AuAu collisions at RHIC by 
the PHOBOS experiment at $\sNN=130$~GeV in 45\%--50\% centrality interval.   
\ct{phobos-all}. The 
 open 
 triangles  show the distributions 
measured
in $\pbp$ interactions
by the UA5 experiment at  the SPS at $\spp=200$~GeV \ct{ua5-zp-rap}.
 The 
 solid 
 squares show the distribution calculated from 
Eq.~(\ref{rapdist}) by using the UA5 $\pbp$  data at 
$\spp \approx  3\,\eNN$ (see Eq.~(\ref{Eeff}) for the definition of 
$\eNN$). 
 The solid circles show  the  beyond-midrapidity part 
 obtained from the calculations using
 the 
energy-balanced 
 limiting fragmentation scaling, i.e. under the shift $\eta
\to \eta-\ln(\eNN/\sNN)$.
 The negative-$\eta$ data points for $\pbp$  interactions are the 
reflections of the measurements taken in the positive-$\eta$ region.   
(b) 
Same  as (a) but the measured distributions of AuAu and $\pbp$ 
collisions  
are 
shifted
by the beam rapidity, $\eta'=\eta-y_{\rm beam}$, with
 $y_{\rm beam}=\ln (\sqrt{s}/m_p)$, where $s$ is, correspondingly, 
$s_{NN}$ or $s_{pp}$, 
 and  
 the calculated distribution is shifted 
  to
  $\eta'=\eta-y_{\rm eff}$
 %, 
 with 
 $y_{\rm eff}= \ln(\eNN/m_p)$.
 The  
distribution measured in AuAu collisions and the calculated distribution 
coincide 
in the 
fragmentation 
region,  when being shifted by  $y_{\rm beam}$ for AuAu data and by 
$y_{\rm eff}$ for the calculations, that 
represents 
the   
 energy-balanced limiting fragmentation scaling.       
\label{fig:rapcentr}}
\end{figure*}

 %\nopar
 %{\bf 5. } 
 \section{Pseudorapidity density distribution in heavy-ion collisions: 
central \vrs\ noncentral collisions}
 \label{sec:rho}
 Figure~\ref{fig:rapmcentr} shows the 
distributions of charged particle 
 pseudorapidity density  
per pair of participants measured 
 in head-on and very central 
 heavy-ion collisions and in 
$pp/\pbp$ interactions.
 The heavy-ion data represent the 
PHOBOS measurements made in 
AuAu collisions  
 at the 
 RHIC at $\sNN=$~19.6, 62.4 and 200 GeV  \ct{phobos-all} and the 
ALICE measurements from  PbPb collisions 
at the LHC at $\sNN=$~2.76 TeV 
\ct{alice-mult}.
 The distributions from $pp/\pbp$ interactions are taken  
as measured by the UA5 experiment 
 \ct{ua5-zp-rap} at $\spp=$~53  at the ISR and at $\spp=$~200 GeV at the 
 SPS,   by the P238 experiment at the SPS \ct{p238} and by the CDF 
experiment 
at 
the Tevatron 
 \ct{cdf630} at $\spp=$~630 GeV, 
 and 
 by the CMS \ct{cmspp7pTn}, LHCb  
 \ct{lhcb7rap} and TOTEM \ct{totem7} experiments at $\spp=$~7 TeV at the 
 LHC. 
 The data shown are taken at the c.m. energies
 $\spp\approx 3\,\sNN$ or $3\,\eNN$. 
 Except for the CMS measurements, 
  the negative-$\eta$ data points from 
$pp/\pbp$ 
interactions are the 
reflections of the measurements taken in the positive-$\eta$ region.

 Within the considered model of constituent quarks and the Gaussian form 
of the 
 pseudorapidity  distribution 
 in Landau hydrodynamics, the relationship between the pseudorapidity 
density distributions $\rho(\eta)$ and $\rho_{pp}(\eta)$ reads 

 \begin{equation} 
\frac{\rho(\eta)}{\rho_{pp}(\eta)}\!=\! 
\frac{2N_{\rm{ch}}}{N_{\rm{part}}\, N_{\rm{ch}}^{pp}} 
 \, \sqrt{1\!+\!\frac{2\ln 3}{L_{N\!N}}} \,
 \exp\! \left[ \frac{-\eta^2}{ L_{N\!N} \left( 2\!+\!L_{N\!N}/\ln 3 
 \right) } \right]\!. 
 \label{rapdist}
 \end{equation}
 Here, all variables
 are defined 
       the same
 %in a 
 way 
 %it is done 
   as 
 in 
Eq.~(\ref{eqn1}), \ie\  taking into account the constituent
 quark scaling of the 
c.m. energy as 
 soon as 
 one relates  $pp/\pbp$ interactions to  central 
heavy-ion 
collisions.

Using Eq.~(\ref{rapdist}), the heavy-ion 
distributions 
 are calculated 
 based on the $\rho_{pp}(\eta$) 
spectra shown in 
Fig.~\ref{fig:rapmcentr}.
 The calculated distributions 
 are shown by solid symbols in 
Fig.~\ref{fig:rapmcentr}.

 One can see that the calculations are in very good agreement with the 
 measurements. Minor deviations are due to some mismatch 
 between $\spp$ and $3 \sNN$ (or $3\, \eNN$) and, 
    as expected,
 due to a slight
 noncentrality;
 this is especially visible at $\sNN=$~19.6 GeV where the energy 
 mismatch  
 is of a largest fraction.  It is 
 %amazing 
 noticeable
 how 
 well 
 the
 PDE
 %energy dissipation 
 picture 
 %of constituent quarks
 allows 
 one
to reproduce the pseudorapidity density 
distributions 
 from heavy-ion interactions  
in the full-$\eta$ range, from central-$\eta$ to forward-$\eta$ regions, 
in the 
 $\sNN$ range spanning 
over more than 2 orders of magnitude.  
 Remarkably,
 the pseudorapidity density distributions, measured in 
$pp/\pbp$ collisions,
 despite
 being
 above those measured in heavy-ion collisions at 19.6 GeV  
 or, on the contrary,
 lying far below the heavy-ion data 
from the LHC
almost in the full-$\eta$ range, 
 equally well reproduce the heavy-ion data as soon as 
 being
 recalculated within the 
 PDE 
  approach.
 % of the dissipating energy of participants. 
 Interestingly, 
 %for 
 the calculations at the LHC energies, 
 well reproduce the heavy-ion data 
 despite
the $pp$ measurements from the three different experiments
 are combined.
 %used. 
  %, which nevertheless 
 %as soon as all the $pp$ 
 %data are combined for the calculations.
 %Some
 A slight deviation in the negative-$\eta$ region is due to some 
asymmetry 
in the ALICE data.

Let us now address 
peripheral collisions 
 to clarify the deviation 
 in centrality dependence between the data and the calculations 
 as it is observed 
 in Fig~\ref{fig:nvsnpart}.

In Fig.~\ref{fig:rapcentr}(a),  
the distribution $\rho(\eta)$ measured \ct{phobos-all} in AuAu collisions 
by the PHOBOS experiment at $\sNN=130$~GeV at 45%--50\% centrality, 
$\alpha=0.475$, is shown along with the $\rho_{pp}(0)$ distribution
 measured in $\pbp$ collisions by the UA5 experiment at $\spp=$~200 GeV 
\ct{ua5-zp-rap}, \ie\ at $\spp\approx 3\, \eNN$
 according to 
 %the approach considered.
 our approach.

Applying Eq.~(\ref{rapdist}), we calculate the $\rho(\eta)$ spectrum which 
is shown in Fig.~\ref{fig:rapcentr}(a) by solid squares. The calculations 
 agree well with 
the measurements in the central-$\eta$ region while fall 
below the 
 data outside this region. This finding 
 shows 
 that 
 in noncentral collisions, the 
 calculations within the approach, which combines the constituent quark 
picture and the relativistic Landau hydrodynamics, reproduces well the 
pseudorapidity density around the midrapidity while 
underestimate the mean multiplicity. The former conclusion is well 
confirmed by our recent studies reported in \ct{edward2} for the 
midrapidity observables, and the latter 
one is demonstrated by Fig.~\ref{fig:nvsnpart}.

To 
 clarify 
 the obtained features, the following 
 comments
 are 
 due.
 %made.

In the PDE picture proposed 
 here, 
 the global observables are defined by the 
energy 
of the participating constituent quarks pumped into the overlapped zone of 
the colliding nuclei. Hence, the bulk production is
 driven by the initial energy 
 deposited
at zero 
 time at  
 rapidity $\eta=0$, similar to the 
Landau hydrodynamics. Then, as 
 %%it 
 is expected and 
 %%commented  just 
 mentioned 
 above, the 
 pseudorapidity density  
  {\it at midrapidity} 
is well 
reproduced for {\it all types} of nuclear collisions, from 
 the 
most central to 
peripheral 
ones. As shown in \ct{edward2}, similarly, the centrality dependence of 
the 
transverse energy density 
at 
midrapidity 
is well reproduced by the calculations and complements 
the 
c.m. energy dependence 
of the 
 head-on data. 
 %within the dissipating effective-energy picture.
 Note that this similarity in 
  the pseudorapidity 
density and 
the transverse energy pseudorapidity density
 is 
 in accordance with the same functional form of the 
(pseudo)rapidity density distribution obtained either in the 
longitudinally 
expanding system considered in the original  
Landau model or 
 when the 
 development 
 in the transverse direction is 
 included
\ct{LandaupTa,LandaupTb,Feinberg}.

From Fig.~\ref{fig:rapcentr}(a), one can see that the 
 calculated distribution $\rho(\eta)$ is narrower than that of the data. 
 The narrowness 
 of the calculated distribution with respect to 
the measured one is explained by a smaller value of $\eNN$ compared to 
 the value of the 
 actual 
 collision
 energy $\sNN$.  
 %while 
 However, 
 the 
calculations in 
Eq.~(\ref{rapdist})
are made with the 
 multiplicity $N_{\rm ch}$ 
taken from the most central nucleus-nucleus 
collisions at  the c.m. energy equal to  $\eNN$
 (and similarly in
 Eqs.~(\ref{pmult}) and (\ref{pmultc}) 
for the midrapidity density $\rho(0)$).  In other words, 
in the
 approach applied 
 here, 
  similar to the Landau hydrodynamics, the
collisions of nuclei   
are 
treated 
 head-on-like.
\\

 %\nopar
 \section{Energy-balanced limiting fragmentation}
 \label{sec:eblf}
 %{\bf 6. } 
 It is established 
that 
at high enough energies,
in different types of interactions 
the 
pseudorapidity density 
 spectra, 
 measured 
at different 
 c.m. energies, become similar in the fragmentation region. 
 %\ie, 
 It means that they 
 are 
 independent of a projectile state (which is the beam or target rest
 frame) for the same type of colliding objects, 
 \ie\ being considered 
as 
a function of 
$\eta'=\eta-y_{\rm beam}$, where
  $y_{\rm beam}=  \ln (\sNN/m_p)$ is the 
beam rapidity
 \ct{book,Landau-difint}.
This observation 
obeys 
  a hypothesis of the  
 limiting fragmentation  
 scaling 
\ct{limfrag}.
  
  Considering
 the 
 limiting fragmentation  
 hypothesis
 %is  applied 
 within the effective-energy 
 approach, 
 one expects the limiting 
fragmentation scaling of the 
distribution $\rho(\eta)$, 
 which is
measured 
 at $\sNN$, 
to be 
similar to 
 that 
 of 
 the calculated distribution 
 but 
   taken 
 at  
 the effective energy 
$\eNN$.
 Note that the limiting fragmentation phenomenon, though 
 being  
 expected as an universal phenomenon for the Gaussian form of 
$\rho(\eta)$ \ct{Landau-difint,wongLH,wongEbE}, 
 naturally 
  arises 
  in Landau hydrodynamics \ct{Landau}.     
 
 In Fig.~\ref{fig:rapcentr}(b), the limiting fragmentation hypothesis is 
 %shown being 
 applied to both the measured and the calculated pseudorapidity 
density distributions $\rho(\eta)$
 from
 Fig. \ref{fig:rapcentr}(a)
using the c.m. energy and 
 the effective energy, respectively. 
Therefore
 the 
measured 
distribution $\rho(\eta)$ is 
shifted by the beam 
rapidity, 
 $y_{\rm beam}$, 
  while
 the calculated distribution from Eq.~(\ref{rapdist})  is shifted by $y_ 
{\rm eff}=
 \ln (\eNN/m_p)$  
and becomes a function of
   $\eta'= \eta - 
 y_{\rm 
 eff}$, as expected.
 One can see that
 the calculated  $\rho(\eta')$ distribution
 of noncentral heavy-ion collisions
agrees 
well with 
 the measured
 distribution $\rho(\eta')$.
 %, as expected.  
 This finding points to a new energy scaling 
 % revealed by using 
 as soon as 
 the 
 %participant 
effective-energy approach
 is applied.  
  In analogy 
with 
the limiting fragmentation scaling, we call 
 the 
 observed  
 scaling 
 the
``{\it energy-balanced}
limiting 
fragmentation 
  scaling''.
  Due to this scaling, the calculated pseudorapidity density is  
getting corrected outside the central-$\eta$ region accordingly.

 To this end, 
  in Fig.~\ref{fig:rapcentr}(a), 
the calculated distribution $\rho(\eta)$ is 
shifted 
by the difference $(y_{\rm eff} -  y_{\rm 
 beam})$
in this 
 region:
$\eta \to \eta-\ln(\eNN/\sNN)$, or, using the effective energy definition,
Eq.~(\ref{Eeff}), 
 $\eta \to \eta-\ln(1-\alpha)$. 
 The calculated distribution  $\rho(\eta)$, where the shift is applied,
 %This shift  of the calculated 
 is 
 shown 
 by the solid circles
in Fig.~\ref{fig:rapcentr}(a). 
 The 
shift
  {\it balances}  the energy
 %adds the needed {\it energy balanced}
 %ingredient to the calculations 
 %providing the description of 
 and this brings 
 the calculations to
the measured 
pseudorapidity density distribution in the full-$\eta$ range in 
noncentral 
 heavy-ion collision.
 It is clear that in 
  head-on  or very central 
collisions, 
 $\alpha$ 
 approaches zero 
 %what 
 which
 makes the shift 
 negligible
 %, in order  to reproduce the data 
 (cf.
Fig.~\ref{fig:rapmcentr}).

 This finding 
 allows 
 %to 
 obtaining 
 $N_{\rm ch}$
 within 
the
 PDE
 approach.
 % of the dissipating 
 % effective-energy of 
 %quark participants. 
  Namely, the difference between the two $N_{\rm ch}$ values, one obtained 
by  
integrating  
 the calculated pseudorapidity density distribution from 
 Eq.~(\ref{rapdist}), and another one of the same distribution 
 but being
  shifted to the left by $\ln(1-\alpha)$, is added to the  
$N_{\rm
ch}$ value obtained from Eq.~(\ref{pmultc}).    
 Where no pseudorapidity density distributions are available in $pp/\pbp$ 
measurements at
$\spp= 3\, \eNN$, the energy-balanced limiting fragmentation scaling is
 applied to reproduce the calculated $\rho(\eta)$: 
 the measured
distribution from  a 
noncentral 
heavy-ion collision
is shifted by $(y_{\rm beam}-y_{\rm eff})$, 
 \ie\ 
 $\eta\to 
 \eta+\ln(1-\alpha)$.
 Then $N_{\rm ch}$ is calculated as above, by adding 
to the calculation of  
  Eq.~(\ref{pmultc})
 the difference between the integral from 
the obtained shifted 
 distribution and the measured multiplicity in this 
noncentral
heavy-ion collision.

 Using
 this 
 ansatz, 
 the values of $N_{\rm ch}$ are calculated for 
 each 
 centrality for the RHIC measurements. 
The 
 calculations 
 are 
 shown by open squares in Fig.~\ref{fig:nvsnpart}.
 One can see that  now  
the calculations 
 well 
reproduce the
measurements from RHIC, 
 with
 no deficit 
in  noncentral collisions.

The energy-balanced limiting fragmentation 
scaling 
 provides an explanation of  the 
``puzzle'' between 
the 
centrality independence 
of the $N_{\rm part}$-normalized mean multiplicity and the 
monotonic 
decrease of 
 the normalized midrapidity pseudorapidity density with the 
centrality, as observed at RHIC. 
 As shown above, the 
pseudorapidity density at midrapidity is determined by the effective 
energy 
of 
centrally colliding
nucleon participants. Hence, the value of this observable increases 
towards 
head-on collisions as soon as the effective energy, 
 made 
available for particle production,
 increases with increasing number of 
participants (decreasing centrality). However, the  multiplicity 
 is measured in the full $\eta$-region, so it
gets additional contribution from beyond the midrapidity. 
 In the context of the PDE picture, this 
 contribution is due to the  balance 
 between  the collision c.m. energy shared by all nucleons of colliding 
nuclei and the centrality-defined 
effective energy of the  interacting participants. The more 
peripheral is the 
collision, the larger the additional contribution is. 
 This contribution 
can 
be directly estimated by the energy-balanced limiting fragmentation 
scaling, 
introduced
here,
which leads to 
 %%the 
 scaling between the 
measured pseudorapidity distribution
and 
 %%the 
distribution calculated 
within the 
 PDE 
 %dissipating participant energy 
 approach.
 % using
 %the effective-energy concept. 

 From Fig.~\ref{fig:nvsnpart}, one can conclude
 that, in contrast to the RHIC measurements,
 almost no  additional contribution
 is needed  for the 
 PDE 
 %participant dissipating energy 
 calculations of 
Eq.~(\ref{pmultc}) in order
to 
describe the 
LHC mean  multiplicity data. As 
the calculations imply, they   
   are made by considering 
the nucleus-nucleus collisions  
 as head-on collisions at the c.m. energy of the value of $\eNN$ 
($\rho(0)$ in Eq.~(\ref{pmultc}) 
as well as
  $N_{\rm ch}$ 
in Eq.~(\ref{rapdist}) are taken from the head-on $\sNN$ fits).  
 %Meantime, 
 However,
 as shown above, the additional contribution to the mean 
multiplicity 
 increases with increasing collision centrality, \ie\ while
going towards more peripheral collisions.
 For head-on collisions, however, this contribution  tends to zero. 
 Given the multiplicity measurements at the LHC are well reproduced 
without the energy-balanced 
 additional contribution,
 one 
concludes that 
 in
  heavy-ions collisions at the LHC  at TeV energies
    the multihadron production obeys a head-on collision regime,
 for all the
 centrality 
 intervals
 measured.
  This points to  apparently different regimes of  
hadroproduction
 occurring
 in heavy-ion 
collisions 
 with 
 $\sNN$ 
 %energies 
 between
 %of 
 a 
 few hundred GeV and TeV energies.
 % and in those happening at TeV energies. 
 This observation supports a similar conclusion 
made above, which is  suggested from the observation of a change of the 
functional type of the fit 
 needed
to describe the energy behavior as soon as the LHC data are included, see 
Fig.~\ref{fig:nvss}. 

 The discussed difference between the mean multiplicity, 
and 
hence the full 
pseudorapidity density distributions, measured in heavy-ion collisions at 
RHIC and at LHC have
 been earlier addressed in \ct{wolschin,wolschin-ln3}, where the 
model of three sources, the gluon-gluon midrapidity and two quark-gluon 
 fragmentation sources, are applied to 
 understand the observations from experimental data.
  In the context of the
 PDE
 approach 
 %of the energy dissipation of constituent quark participants 
 given here, the difference in the nature of collisions at 
effective c.m. energy is appealed to
 explain different centrality dependence of the data from the two 
colliders.  
 Similarly to calculations in \ct{wolschin,wolschin-ln3}, additional 
contribution from 
the fragmentation regions are shown to be needed at RHIC. 
 %Meantime, 
 However,
 no  
such contribution  is needed at the LHC energy. 
 Meantime, the midrapidity pseudorapidity densities measured at RHIC and 
 at LHC
 %, 
 % however, 
 do not show different 
 behavior with centrality 
 %from RHIC to LHC energies 
 and are found \ct{edward2} to be similarly well reproduced by the 
 PDE 
 %effective-energy dissipation 
 calculations 
 %considering head-on collisions 
 %at effective c.m. energies, so that 
 where
 no preference 
 %to be 
 is
 given 
to midrapidity 
or 
fragmentation sources.

 There are 
 other approaches, which also consider the three effective regions in 
pseudorapidity density distributions of charged particles produced in 
$pp/\pbp$ and in heavy-ion collisions. 
 In the string percolation model 
\ct{percol1,percol2}, the fragmentation region is 
populated by 
strings of valence quarks and the midrapidity region by additional short 
strings between quarks and antiquarks. 
 %Within these 
 In other approaches,  one 
introduces a leading 
particle
activity 
within the hydrodynamic \ct{hydro-lp1,hydro-lp2,hydro-lp3,liu-3source} or 
thermal 
\ct{therm-lp} 
pictures of 
 the
 multiparticle 
 production 
 processes.
 %Al
 Like in the 
 PDE
 %energy dissipation 
 consideration, 
 presented here, a 
similarity of the mechanism of 
particle 
production   in
$pp/\pbp$ and heavy-ion collisions is 
 also
 assumed in these approaches.
 % However, 
 Whereas
 within the 
 PDE
approach
 %of the 
 %dissipating 
 %energy 
 %of participants
the leading particles resulting from the 
spectators 
 are considered 
 to be produced
in nucleon-nucleon 
collisions, where 
 a single quark pair interaction  
 is assumed,  
no leading particle effect is implied 
for central nucleus-nucleus collisions,
 where
 the entire 
 energy of the participants is considered to be available for bulk  
 hadron
 production.
 As already noticed above, also no difference in the particle production 
sources in different pseudorapidity regions is assumed in the PDE 
approach.
 %,  resulting, however, in 
 %successful  description of the energy and centrality 
 %dependencies of the multiplicity as well 
 %as the pseudorapidity distribution relating  $pp$ 
 %and heavy-ion collisions.
 Then, the c.m. energy scaling due to the key 
picture of the constituent quarks, applied to the Landau hydrodynamics, 
 allows 
 revealing the universality of   
 the multihadron dynamics in hadronic and nuclear interactions.
 \\

\section{Multiplicity effective energy dependence in head-on 
and noncentral 
collisions}
 \label{sec:nvseeff}
 %{\bf 7. } 
Given the obtained agreement between the data and the 
 calculations,
 and 
 considering the similarity put forward for  $\eNN$ and 
$\sNN$, one would expect the measured 
centrality data at $\eNN$ to follow the  $\sNN$ dependence of the 
mean multiplicity in the most central nuclear collisions. In 
Fig.~\ref{fig:nvss}, the measurements of the charged particle 
mean multiplicity of 
 {\it head-on} 
 nuclear collisions are added by the 
 {\it centrality} measurements 
 by the PHOBOS 
\ct{phobos-all} and the 
ALICE \ct{alice-mult2} experiments
(Fig.~\ref{fig:nvsnpart}) 
 where the 
 centrality data 
 are plotted  
as a 
 function of $\eNN$.
 Due to the above finding of the energy-balanced  
 limiting fragmentation scaling,
 explaining
 the 
 lack of 
centrality dependence of the 
mean
multiplicity at RHIC energies, these data are
 plotted by 
subtracting the energy-balanced contribution. In addition,   
 the  centrality data at $\sNN=$~19.6 GeV  are 
 %given while are not 
 shown 
in 
Fig.~\ref{fig:nvss}  but not in 
Fig.~\ref{fig:nvsnpart}. 
 From 
Fig.~\ref{fig:nvss}, 
one concludes that 
effective-energy dependence 
of  the centrality data 
 complements 
 the c.m. energy behavior
 of the 
 head-on 
 collision 
data.

 To better trace the similarity 
between the head-on 
collision and centrality data, 
 we fit the $\eNN$-dependence 
 of 
the centrality data 
by 
the hybrid and the power-law functions,
 similarly to the head-on collisions.
 For the hybrid fit one gets

\begin{eqnarray}
\nonumber
 \frac{2\, N_{\rm ch}}{N_{\rm part}} =
(3.04\pm 0.60)-(1.40\pm 0.24)\ln(\eNN) 
& \\ 
+(1.12\pm 0.04)\ln^2(\eNN)+(0.032\pm 0.028) \eNN^{(0.848\pm 0.106)}.
&
\label{hybnaac} 
\end{eqnarray}
 This fit is shown by the dashed line in 
Fig.~\ref{fig:nvss}.  This fit agrees well with the same type of the fit 
to  the
head-on 
collision 
data in the entire available energy range though lying slightly 
above the latter one for the data at 
$\sNN \lssim 10$~GeV.  
 For the log$^3(\eNN)$ fit function of the three-sources approach, similar 
to 
Eq.~(\ref{ln3}), one 
finds:
 % \begin{equation}
 \begin{eqnarray}
\nonumber
 \frac{2\, N_{\rm ch}}{N_{\rm part}} 
= 
 (1.70\pm 1.49)  + (1.18\pm 0.54)\, \ln(\eNN) & \\ +(0.152\pm 
0.008)\,\ln^3(\eNN). &
\label{ln3c} 
 \end{eqnarray}
 %\end{equation}
The fit is shown by the thin dashed-dotted line in 
Fig.~\ref{fig:nvss},  
and lies on top of the 
analogous fit, 
Eq.~(\ref{ln3})
to the head-on data,  
 except a slight 
enhancement 
at $\sNN \lssim 10$~GeV, similar to the hybrid fit. 
The power-law $\eNN$-fit
 for the centrality data is
shown by the 
dotted 
line  
in 
Fig.~\ref{fig:nvss}. It 
 is  found to be 
 similar to the power-law $\sNN$-fit,  Eq.~(\ref{expnaa}), to the 
head-on collision 
data shown by the dashed-dotted line.

From this one concludes that 
within 
the 
 picture proposed here
 the data 
are well 
reproduced 
under the assumption of the effective energy 
 which governs
 %deriving 
 the 
 multiparticle production.
 % process and pointing
 This points  
 to the 
 the same energy behavior   
 in multihadron production for 
all types of heavy-ion collisions, from peripheral to the most central 
 collisions.
 % This observation is similar to that obtained 
 % by us  earlier \ct{edward2} for the midrapidity pseudorapidity 
 % density 
 % and the transverse energy density data at midrapidity.

Here, let us stress an important corollary of the 
 PDE
 % approach 
 %of  the 
 %dissipating effective-energy 
 %of participants
approach. As soon as the  effective energy in nucleus-nucleus 
collisions determines the pseudorapidity density at midrapidity, then 
 the midrapidity pseudorapidity densities at the same effective energy 
but at different c.m. energy get the same value. In other words, the 
densities 
are defined by the effective energy independent of the energy of the 
collision.
 % This 
 %  was indeed observed 
 %  in \ct{edward2} using RHIC data. 
 The observation made here for the 
multiplicity dependence on the effective energy confirms 
 the 
observation
 made earlier for the midrapidity densities \ct{edward2}, 
while 
 it 
 adds another important ingredient which takes into account the 
additional energy-balanced
contribution to the mean multiplicity in noncentral nucleus-nucleus 
collisions.

   From the  hybrid fits obtained, we estimate the multiplicity for the 
future LHC
heavy-ion run. 
 Since the 
hybrid fit for 
 the
head-on collision data
and the fit to the centrality data
 show slightly different  increase 
 with c.m. energy, 
the predictions of the two fits are averaged. Hence, the 
mean multiplicity $2N_{\rm ch}/N_{\rm part}$  
value is predicted to  be  about 
119  
with 5\% uncertainty in the 
most central heavy-ion collisions at $\sNN=5.13$~TeV.
 The prediction is  shown by 
 the 
 right-inclined hatched area 
in Fig.~\ref{fig:nvss}. 
This value is close to  the value of about 116 one gets from the ALICE fit 
\ct{alice-mult} and 
about 108 which one obtains from Eq.~(\ref{ln3}).
In addition, the fit-averaged 
prediction based on
  $pp$ collisions at $\spp=13$~TeV,  
 %being 
 recalculated within the
 PDE 
 approach, 
 %of the dissipating energy 
 %of participants, 
 is shown 
in Fig.~\ref{fig:nvss} as the left-inclined hatched area.
 
 The predictions are made as well for the centrality dependence and are 
given in 
Fig.~\ref{fig:nvsnpart}.  We give the predictions for $\sNN=5.02$~TeV, 
considering the recent measurements reported  by ALICE for the rapidity 
density \ct{alice5020} and expecting the mean multiplicity measurements 
 at this energy. 
 %The predictions should not differ much from  those at 
 %$\sNN=5.13$~TeV.
 The two 
types of predictions are shown.
  
 First, similar %%ly 
 to the above predictions made to the head-on collisions, 
we 
use the fit functions. As soon as, within the effective-energy approach, 
we 
treat noncentral collisions as central collisions at energy $\eNN$, 
then we 
use 
the head-on collision multiplicity fits to predict the centrality 
dependence. This is similar to that made for the existing data as shown in 
Fig.~\ref{fig:nvsnpart} by the dashed lines. However, for  the 
predictions, 
we use the average values of the hybrid and the power-law fits, 
Eqs.~(\ref{hybnaa}) and (\ref{expnaa}), as soon as those 
deviate for $\sNN$ above 
2.76~TeV. The prediction for 
$\sNN = 5.13$~TeV
centrality dependence is shown by the 
dashed-dotted line. The 
 centrality and $N_{\rm part}$ 
values are
alike in the 2.76 TeV data shown. The expectations show
 an
increase of the mean multiplicity with $N_{\rm part}$  (decrease with 
centrality)
from about 52 to about 118. The increase looks to be slightly faster than 
at
$\sNN = 2.76$~TeV, especially for the peripheral region.

 Second, the PDE set of prediction is made using the calculations 
based
 on Eq.~(\ref{pmultc})  combining the 
constituent quark model and the Landau hydrodynamics. 
 This prediction for the centrality dependence is shown by the 
solid inverted triangles. One can see that the predictions are close to 
ones 
obtained from the head-on collision data fits. 

The predictions are compared with the LHC data at 
$\sNN=2.76$~TeV. 
 To better match the predictions for highly central collisions, the 2.76 
TeV data points are  multiplied by  1.3.
 One can see  that 
 the predictions   are well reproduced by 
 the scaled data. This indicates no change of the hadroproduction 
mechanism expected with increase of the c.m. energy at LHC, in contrast to 
what is seen 
as one compares the scaled  200 GeV RHIC data with the 2.76 TeV 
measurements. 

 An interesting issue to be addressed in the framework of the PDE picture 
is 
 %%to consider 
 asymmetric collisions, such as nucleon-nucleus 
($p/d$-nucleus) 
ones. In these  interactions the multiplicity is also expected to have 
no centrality dependence. This is  due to the many nucleon-nucleon 
interactions of the 
incident proton with  the nucleons of the interacting 
nucleus while  the secondary particles
produced in the reaction are assumed to be created out of the c.m. energy
deposited to the interaction zone. 
 The proton and nucleus are considered to interact via a single 
pair of 
constituent
quarks, one from the proton and another one from a nucleon
in the interacting nucleus. Then, no centrality dependence of the 
multiplicity is expected 
in $p$-nucleus collisions  with the multiplicity values  to 
be 
similar to 
that   from $pp/\pbp$ interactions at the c.m. energy $\spp
\simeq 
\sNN$.
  As a consequence, this, at a given centrality,  results  
 %%to
 in 
 $N_{\rm 
part}/2$  for  the ratio  $N_{\rm ch}/N_{\rm ch}^{pp}$.
  These features have been indeed obtained in $d$-Au interactions 
at  RHIC \ct{phobos-dAu}. 
 Moreover, the effect of the 
$N_{\rm part}$-dependence  
of the multiplicity ratio 
  obtained at RHIC
 %is shown
 %to be true 
 has been also observed 
 % also for 
 in
hadron-nucleus collisions
at lower
$\sNN \approx$~10--20~GeV \ct{phobos-dAu,busza}.
 These observations seem to be also obtained 
at 
 LHC, where 
the 
dependence of
the pseudorapidity density on $\sNN$ measured in   $p$-Pb collisions 
at $\sNN= 
 5.02$~TeV is observed \ct{alice-pPb} 
 to be consistent with a power-law $\spp$-fit to 
 %inelastic 
 $pp/\pbp$ data 
and its centrality dependence is 
shown~\ct{alice-pPb-centrality,atlas-pPb-centrality} to 
 demonstrate the
importance 
of nucleon-nucleon interactions for 
 %understanding  
 $p$-Pb results. 
 \\

 \section{Summary and Conclusions}
 \label{sec:sum}
 %\nopar 
 %{\bf 8. } 
 In summary, the multihadron production process in nucleus-nucleus 
 collisions  
 and its 
  universality in nuclear and hadronic interactions 
 are  
 studied. The study exploits 
 %using 
 the charged particle mean multiplicity 
 dependencies on the c.m. collision energy per nucleon,  $\sNN$, and on 
the 
number of nucleon participants, or centrality, measured 
in the 
 energy range of a few GeV to a few TeV.
 The study is carried out in the framework of the earlier proposed  
approach
of the dissipating effective energy 
 of constituent quark participants 
\ct{edward,edwarda}, or the participant dissipating energy (PDE) 
approach. In this 
approach, the
participants are considered to 
 form the initial zone of a collision and to determine the production of  
hadrons at the very early stage of the collision. 
 In this consideration one  combines the constituent quark picture with 
Landau hydrodynamics and interrelates the multihadron production in 
different types of collisions by
 a
 proper scaling of the c.m. energy of 
collisions. 
 In particular, an energy-scaling factor of 1/3 in $pp/\pbp$ 
measurements is shown to reveal the universality of the multiplicity 
dependencies in nucleon-nucleon and 
nucleus-nucleus interactions.  

 %  Recently, the c.m. energy dependence of the 
 % midrapidity pseudorapidity 
 % density and the transverse energy density at midrapidity 
 % measured 
 % in the head-on nucleus-nucleus collisions have been shown to be 
 % well  described using this approach \ct{edward2}.  
 % Moreover, the c.m. energy dependence of the studied variables 
 % as measured  in head-on collisions has been 
 % shown to be consistent with the energy 
 % dependence of the centrality data as soon as the effective energy 
 % term is introduced.
 % Then, the centrality dependence of these two variables 
 % has been demonstrated to be well described by the calculations 
 % within the  PDE approach.
 %% of the dissipating  energy 
 %% of participants. 

 % In this paper, the energy and centrality dependences are studied 
 % for the mean multiplicity 
 % extending the earlier energy-dependence analysis \ct{edward,edwarda} 
 % above  the RHIC energies 
 % and adding to that the centrality dependence study.
 In 
 the entire available  $\sNN$ range of about a few TeV,
  the energy dependence of the multiplicity in 
head-on collisions is found to be well described by the 
calculations
 performed 
 within the effective-energy approach
based on $pp/\pbp$ data. 
  Meanwhile, depending on the data sample,
the calculations
are found either to describe the 
measured centrality
dependence 
or to show some deviation 
between 
the 
calculations and 
the 
data. 
For the
RHIC data,  
the deficit in 
the predictions 
is observed for noncentral collisions so that the 
predictions do not follow a 
 constancy  with the centrality as it is  observed  at RHIC.
 %In contrary, t
 The LHC mean multiplicity centrality dependence, however,
 is found to 
be 
well described by the  
 calculations 
 %: both the data and the calculations  
 %showing an
   including the 
 increase towards the most 
central collisions.    
  
 To clarify  the observations, 
approach 
of the  effective-energy 
of the quark participants 
is applied to 
 the pseudorapidity density distribution measured in heavy-ion collisions.
 The 
 energy-balanced limiting fragmentation scaling is introduced 
 based on 
 %under 
 assumption of the similarity of the fragmentation region of the 
measured 
distribution in the beam rest frame and that 
 determined 
 from the calculations by using 
 the effective energy.
 The revealed scaling 
 allows 
 us
 to reproduce the pseudorapidity 
density distributions 
independently
  of the centrality of collisions and then to correctly 
describe 
 the centrality independence of the  mean multiplicity 
 measured at RHIC.
 Moreover, 
 this finding provides a solution to the RHIC ``puzzle'' of the 
 difference 
 between the centrality 
independence of the  mean multiplicity  \vrs\ the monotonic 
decrease of the midrapidity pseudorapidity density with the 
increase of centrality.
 The  mean multiplicity is shown to get a fraction of additional 
contribution 
 to account for 
 the 
 balance 
 between  
 the collision c.m. energy shared by all nucleons 
 and 
 the effective 
 energy of the participants.
 %Meantime, 
 However,the midrapidity pseudorapidity density is fully defined by the 
effective 
energy of colliding participants.

 Given 
that 
the calculations made in the context of the proposed approach are 
 considering central collisions of 
nuclei,  an agreement between the calculations and the LHC data
indicates that at TeV energies 
 the collisions seem to present  head-on collisions of the participants at 
the c.m. energy 
 of the 
 scale
 of 
 the effective 
energy. 
 %Then, 
 Thus,
no energy-balanced 
additional contribution
is required even 
 with a 
 %%at 
 relatively small number of 
participants at TeV energies.

 Based on the above findings, the complementarity of the head-on 
collisions and the centrality data is shown resulting in the similar 
energy behavior of the mean multiplicity measurements as soon as the 
data are considered in terms of the   
effective energy.  
 A departure  of the c.m. energy dependence
 of the data from the second-order logarithmic behavior to the power-law 
or higher-order logarithmic polynomial function one observes 
at  
 the region of 
 1 TeV
 suggests a 
    transition  to  a new 
regime in nucleus-nucleus collisions at TeV energies.
 Interestingly, these findings made for a full collision rapidity range
 are  
similar to those  drawn from the studies \ct{edward2} of the 
pseudorapidity density and the  transverse energy density at 
midrapidity. 
 This is also in accordance with the change of the 
multiplicity dependence on centrality which also indicates a 
possible change of 
the regime of multihadron production 
 as one moves from the RHIC to LHC 
energies.

The hybrid and the power-law fits are found to describe well the 
existing 
data on the 
 charged particle 
multiplicity from   
$pp/\pbp$ interactions in the entire c.m. energy $\spp$ range up to the 
top 
Tevatron 
energy of 1.8 TeV. However, in this case no clear change 
from the power-law  behavior to the quadratic log polynomial one  is 
obtained 
in  
the multiplicity c.m. energy 
dependence. Moreover, the 
  predictions made here for the 
mean 
 multiplicity  for $\spp$ in the LHC energy range of 2.36
 %%~TeV 
 to 13~TeV   
within the 
 PDE
 approach
 %.  
 %dissipating energy 
 %of participants 
 demonstrate 
a closeness between the predicted values and 
the lower-energy $\spp$ fit.
   One concludes that, in contrast to heavy ions, 
 no change 
 in  multihadron production in $pp$ 
collisions is 
expected up to the foreseen LHC energy.  

 Based on the results of the hybrid fits, 
 the predictions for the charged particle 
 mean multiplicity in head-on 
heavy-ion collisions 
at 
$\sNN=$~5.13 TeV 
 at the LHC are 
 given.
  Within the obtained complementarity of 
 head-on collisions and 
centrality data, the 
predictions  are made for 
 the mean multiplicity centrality dependence to be measured.

 The soon-to-come measurements at the LHC are of crucial importance for
 further understanding of    the multihadron  
 dynamics. This will 
 shed the light on  
 %and its
 the 
 universality  of the multihadron production process in different
types of
collisions  
 and
 clarify 
 the 
 PDE
 approach 
 %of the dissipating effective-energy of  participants 
and 
 the obtained 
 energy-balanced 
limiting 
 fragmentation which have been 
 shown to  
 successfully 
 describe 
 the features of global key  
 observables  
 by
 relating
hadronic 
  and 
 nuclear 
collisions.
\medskip
\medskip

 %\begin{acknowledgement}
 %{\bf Acknowledgements}
Thanks go to  Lev Kheyn for 
 his interest in the 
 work
 and 
enlightening 
 discussions.
 We are grateful to Gideon Alexander, Amarjeet Nayak, David Plane, 
Vladimir Roinishvili and  
Marek Ta\v{s}evsk\'y  
for their 
help 
during the work on this paper.
 % preparation of the manuscript. 
 The work of Alexander Sakharov is partially supported by the US National
 Science Foundation under Grants No.PHY-1205376 and No.PHY-1402964.
 %\end{acknowledgement}

\end{document}